\documentclass[preprint,12pt]{elsarticle}

\usepackage{amsfonts}
\usepackage{amssymb}

\usepackage{comment}
\usepackage{graphicx}
\usepackage{subfig}
\usepackage{threeparttable}

\journal{}

\begin{document}

\begin{frontmatter}



\title{Performance of Local Information Based Link Prediction: A Sampling Perspective}


\author{Jichang Zhao, Xu Feng, Li Dong, Xiao Liang and Ke Xu$^\star$}

\address{State Key Laboratory of Software Development Environment, Beihang University
\\ $^\star$ Corresponding author: kexu@nlsde.buaa.edu.cn}

\begin{abstract}
Link prediction is pervasively employed to uncover the missing links
in the snapshots of real-world networks, which are usually obtained
from kinds of sampling methods. Contrarily, in the previous
literature, in order to evaluate the performance of the prediction,
the known edges in the sampled snapshot are divided into the
training set and the probe set randomly, without considering the
diverse sampling approaches beyond. However, different sampling
methods might lead to different missing links, especially for the
biased ones. For this reason, random partition based evaluation of
performance is no longer convincing if we take the sampling method
into account. Hence, in this paper, aim at filling this void, we try
to reevaluate the performance of local information based link
predictions through sampling methods governed division of the
training set and the probe set. It is interesting that we find for
different sampling methods, each prediction approach performs
unevenly. Moreover, most of these predictions perform weakly when
the sampling method is biased, which indicates that the performance
of these methods is overestimated in the prior works.
\end{abstract}


\end{frontmatter}


\section{Introduction}
\label{sec:introduction}

Complex networks has been tremendously utilized to characterize the
real world. For instance, the Internet could be treated as a network
constituted by routes and physical links among them. While regarding
to the Facebook, the users could be nodes and the online friendships
could be links, which also compose a network. Hence, the complex
network is a powerful tool to represent the objects and their
relations. Moreover, in the real world, the size of the network
might be extremely large. Taking Facebook as an example, it contains
nearly 600 millions users
currently~\footnote{http://en.wikipedia.org/wiki/Facebook}. Because
of the large scale, it is indeed much hard for the research
community to get a complete and rich picture of the network. In
addition, even for some small ones, it is also difficult to observe
some links in the experiments~\cite{local_bayes}. For the above
reasons, many research interests have been devoted to the problem of
link prediction in recent years. Based on simple local information
or global evolving rules, link prediction can uncover the missing
links in the incomplete network and even predict the future links
that would be generated later.

In the previous literature, in order to validate the performance of
the prediction methods, the edges of the known network is usually
divided into the training set and the probe set randomly. However,
in the real world, sampling a large-scale network is often not pure
random but biased. An intuitive example is the
Breadth-First-Search(BFS) sampling, which is always employed to
crawl the online social networks~\cite{mislove_sns_bfs,
ahn_sns_bfs}. It has been unrevealed that the BFS sampling is not
random but biased to the nodes with higher
degrees~\cite{mhrw_facebook}, which means it would only extract a
dense region of the network without reaching to the other parts.
Then a natural question could be presented that we obtain the
training set and the probe set through random selection is not
reasonable, particular for the snapshot that sampled from the
network by certain biased methods. Intuitively, for a snapshot
obtained from a biased sampling method, the previous evaluation of
the performance for each link prediction method is not convincing
because it just corresponding to the random sampling. Hence, it
becomes difficult for us to select a proper prediction method for a
certain data set if we only rely on the precision validation
obtained from randomly selected probe set. In fact, recent
work~\cite{cold_ends} has also pointed out this problem and given an
excellent illustration through selecting probe set based on the
edge's popularity. However, a complete investigation of how sampling
methods affect the performance of the existing link prediction
approaches still remains unclear. In order to fill this gap, in the
present work, we try to reveal the interaction between sampling
methods and prediction approaches.

We employ five pervasively used sampling methods to generate the
training set and the probe set from nine real-world networks of
different contexts. Given the practical usage, we only consider the
local information based link prediction methods in the present
paper. Through comparison of the performance, we find that for each
of the ten prediction measures, it performs unevenly on different
sampling approaches. Besides, these measures perform poorly as
compared to the case of pure random selection of the probe set,
which indicates that their performance might be overestimated
conventionally.

The rest of the paper is organized as fellows. In
Section~\ref{sec:relatedworks}, we would present the recent related
works. The local information based link prediction and the sampling
methods would be illustrated in Section~\ref{sec:premilinaries}. In
Section~\ref{sec:datasets}, the data sets used in the experiments
are depicted. Then the observations and remarks of the performance
reevaluation in the view of sampling methods are introduced in
Section~\ref{sec:results}. Finally, in Section~\ref{sec:conclusion},
we conclude this work briefly.

\section{Related Works}
\label{sec:relatedworks}

Recent years have witnessed growing interests in the link prediction
of complex networks. It aims to evaluate the likelihood of a link
between two nodes not connected currently, based on the existing
links information~\cite{physica}. The existing methods for link
prediction can be divided into three categories~\cite{xufeng_l}. The
first method defines a measure of proximity or similarity between
two nodes in the network, taking into account that links between
more similar nodes are of higher existing likelihood. Liben-Nowell
and Kleinberg summarize many similarity measures based on node
neighborhoods, the ensemble of all paths and higher-level
approaches~\cite{Nowell}. Motivated by the resource allocation
process taking place in networks, Zhou et al. review the existing
similarity measures and propose a new similarity measure, which has
great performance in several representative networks drawn from
different fields~\cite{RA}. In the present work, we mainly focus on
this kind of methods for their simpleness and efficiency. The other
two kinds of  methods are based on the maximum likelihood
estimation~\cite{Nature} and  machine learning
techniques~\cite{Event_network}. However, in the prior works, the
probe set is generally determined by random edge selection. In the
recent work~\cite{cold_ends}, the authors argue that this
conventional evaluating methods may lead to terrible bias and then
study how to uncover missing links with low-degree nodes.

With respect to the growing networks and their tremendously large
scales, many different sampling methods have been presented in
recent years. The aim of a sampling method is to derive a
representative snapshot of the network with low cost. The simplest
BFS is usually employed to crawl the online social networking sites
and collect online social networks~\cite{ahn_sns_bfs,
mislove_sns_bfs}. Because of the BFS-introduced bias, Gjoka and et.
al consider the Metropolis-Hastings Random Walk, which is first
presented to sample the peer-to-peer network
unbiasedly~\cite{mhrw_p2p}, in crawling the Facebook and achieve the
goal of uniform stationary distribution of
nodes~\cite{mhrw_facebook}. Leskovec and et al. review many sampling
method and present a new approach named Forest-Fire(FF) which
matches very accurately both static as well as evolutionary graph
patterns, while the sampled size decreases down to about 15\% of the
original graph~\cite{ff}. Regarding to the drawbacks of random
walks, a multidimensional random walk,named Frontier Sampling, is
presented in~\cite{fs} and the authors find that this approach is
more suitable to sample the tail of the degree distribution of the
graph. In recent work~\cite{estimate_size}, a new method of
estimating the original network's size is depicted, while in
~\cite{diffusion_ff}, the relation between sampling methods and
information diffusion in social media is also discussed. However, to
our best knowledge, little attention has been paid to employing
these methods to evaluate the performance of the link prediction in
complex networks.

It is worthy to be noted that the recent work~\cite{cold_ends} has
pointed out the problem induced by selecting edges into the probe
set randomly. It also gives a simple but clear illustration of this
problem by use of the edge popularity. However, the interplay
between sampling methods and link prediction approaches is still not
well investigated. Hence, we try to fill this void in the present
work.

\section{Preliminaries}
\label{sec:premilinaries}

In this section, we mainly depict the basic definitions about the
network, then the local information based prediction methods and the
sampling approaches employed later would also be introduced,
respectively.

\subsection{Definitions}
\label{subsec:definitions}

In this paper, all the data sets we use could be denoted as an
undirected graph $G(V,E)$, where $V$ is the set of objects(nodes)
and $E$ is the set of relationships(links) among these objects. For
each node $i$, the number of the links connected to it is defined as
its degree $k_i$, then the averaged degree of $G$ is
$$\langle k\rangle=\frac{2|E|}{|V|}.$$ The nodes that connected to
$i$ is defined as $n(i)$, i.e, $i$'s neighbors. Heterogeneity of the
network, defined as
\begin{equation}
H=\frac{\langle k^2 \rangle}{\langle k \rangle^2},
\end{equation}
is usually used to characterize the nonuniformity of
degrees~\cite{RA}. Clustering coefficient of a node $i$ is used to
characterize how closely its neighbors are connected. It can be
defined as
\begin{equation}
C_i=\frac{2|E_i|}{k_i(k_i-1)},
\end{equation}
where $E_i$ is the set of ties between $i$'s neighbors and $k_i$ is
the degree of $i$. For the case of $k_i=1$, we set $C_i=0$ in this
paper. The averaged clustering coefficient of the network can be
defined as
\begin{equation}
C=\frac{\sum_{\{i \in V\}}{C_i}}{|V|}.
\end{equation} For an undirected edge $e_{ij}$ between $i$ and $j$, we could
define its popularity~\cite{cold_ends} as
\begin{equation}
e_{pub}(i,j)=(k_i-1)(k_j-1).
\end{equation}
Similarly ,we can define the number of the common neighbors for its
two ends as
\begin{equation}
e_{CN}(i,j)=|n(i)\cap n(j)|.
\end{equation}

\subsection{Local information based link prediction}
\label{subsec:linkprediction}

Many link prediction methods has been presented recently. Given the
practical feasibility, we only investigate the local information
based ones for their simpleness and low cost. In fact, for an
arbitrary prediction method, the essence of its algorithm is to
score the node pair $\langle i,j\rangle$, where $i,j\in V$ and are
not connected by an edge in $E$. After allocating a value $s(i,j)$
for each pair $\langle i,j \rangle$, we only need to sort them in
the decreasing order of the score and choose the ones with higher
values as the predictions. We mainly utilize ten popular measures of
$s(i,j)$ in this paper and all of them are introduced in detail as
follows.

\emph{Common Neighbors(CN)} It is assumed that two nodes with more
common nodes are easily to be connected. Then the score of this
methods could be defined as
\begin{equation}
s^{CN}(i,j)=|n(i) \cap n(j)|
\end{equation}
intuitively.

\emph{Adamic-Adar Index(AA)} This method try to assign more weights
to the neighbors with lower degrees~\cite{AA}, hence, the score
could be denoted as
\begin{equation}
s^{AA}(i,j)=\sum_{q\in n(i) \cap n(j)}{\frac{1}{\log{k_q}}}.
\end{equation}

\emph{Resource Allocation(RA)} In this measure, the score between
$i$ and $j$ is defined as the amount of resource that $j$ receives
from $i$~\cite{RA}. Then the score could be denoted as
\begin{equation}
s^{RA}(i,j)=\sum_{q\in n(i) \cap n(j)}{\frac{1}{k_q}}.
\end{equation}

\emph{Salton Index(SAI)} The score is defined as
\begin{equation}
s^{SAI}(i,j)=\frac{|n(i) \cap n(j)|}{\sqrt{k_ik_j}}
\end{equation}
in this measure.

\emph{Jaccard Index(JI)} The score of each pair could also be
obtained from Jaccard's definition as
\begin{equation}
s^{JI}(i,j)=\frac{|n(i)\cap n(j)|}{|n(i)\cup n(j)|}.
\end{equation}

\emph{S$\phi$rensen Index(SPI)} This measure is presented to the
ecological community data sets~\cite{RA}, the score is defined as
\begin{equation}
s^{SPI}(i,j)=\frac{2|n(i)\cap n(j)|}{k_i+k_j}.
\end{equation}

\emph{Hub Promoted Index(HPI)} The score in this method is defined
as~\cite{hpi}
\begin{equation}
s^{HPI}(i,j)=\frac{|n(i) \cap n(j)|}{\min\{k_i,k_j\}}.
\end{equation}
It could be obtained easily that for the links connected to the
nodes with higher degrees(hubs) would be allocated higher
values~\cite{RA}.

\emph{Hub Depressed Index(HDI)} Different from the measure of
\emph{HPI}, in this method, the score is defined as
\begin{equation}
s^{HDI}(i,j)=\frac{|n(i)\cap n(j)|}{\max\{k_i,k_j\}}
\end{equation}
to decrease the values that allocated to the links connected to the
hubs.

\emph{Leicht-Holme-Newman Index(LHN)} This measure is presented
in~\cite{lhn} and it defines the score as
\begin{equation}
s^{LHN}(i,j)=\frac{|n(i)\cap n(j)|}{k_ik_j}.
\end{equation}

\emph{Preferential Attachment(PA)} This measure is motivated by the
mechanism of preferential attachment in the evolution of scale-free
networks and the score could be defined as~\cite{pa}
\begin{equation}
s^{PA}(i,j)=k_ik_j.
\end{equation}
As stated in~\cite{RA}, this method needs minimal information and
computation complexity among all the measures mentioned here.

In summary, all these measures would be employed to score each pair
of unconnected nodes then to predict the pairs that with the higher
scores as the most plausible ones.

In order to evaluate the performance of these methods, in the
previous works, a generally used way is to divide $E$ into two
non-overlapped parts, including $E^{T}$, which is stated as the
training set, and $E^{P}$, which is stated as the probe set or the
testing set. Clearly we have $E=E^{T}\cup E^{P}$ and $E^{T}\cap
E^{P}=\emptyset$. In a general way, $E^{T}$ contains 90\% edges in
$E$ and the remains are allocated to $E^{P}$. For all the possible
links in $\bar{E}$, the prediction methods only use the information
contained in $E^{T}$ to score these links, then sort them in
decreasing order of scores and select the top $|E^{P}|$ ones into
the prediction set $E'^{P}$. Hence, the precision of a prediction
method $\pi$ could be defined as $$Precision(\pi)=|E^{P} \cap
E'^{P}|/|E^{P}|$$ intuitively.

In addition, another pervasively used evaluating measure is AUC,
which could be interpreted as the probability that a randomly chosen
missing link is given a higher score than a randomly chosen
nonexistent link in the prediction method~\cite{RA}. In the
implementation, among $n$ times of independent comparisons, if there
are $n'$ times that the missing link having higher score and $n''$
times the missing link and the nonexistent link having the same
score, then the accuracy could be defined as
\begin{equation}
AUC=\frac{n'+0.5n''}{n}.
\end{equation}
As stated in~\cite{RA}, the extent to which the AUC exceeds 0.5
indicates how much better the prediction method utilized performs
than the random case. In the experiments later, we mainly employ
this measure.

\subsection{Sampling methods}
\label{subsec:samplingmethods}

With respect to the growth of real-world networks, especially the
emergence of the large-scale online social networks, many sampling
methods have been developed to get a representative view of original
network. Inspired by the motivation of this paper, we aims to sample
$s_f$($0\leq s_f \leq 1$) edges from $E$ to obtain $E^T$ and then
the remaining edges could compose $E^P$. Hence, different from the
conventional random selection, the training set and the probe set
are determined by a certain sampling method. In this paper, we
mainly employ five typical sampling methods, which are depicted as
follows.

\emph{Breadth First Search(BFS)} For its simpleness, BFS is always
used to sample the network, especially for crawling the web and
obtaining the online social networks. However, as we have stated in
the former section that BFS is biased to the nodes with higher
degrees and might only extract one dense core of the network without
reaching out to the other parts of the network. The procedure of
this method could be listed as
\begin{itemize}
\item
    Step 1: set all the nodes' states to 0 and randomly select a starting node $i$ from $V$.
\item
    Step 2: add $\forall e_{ij}\in E$ to $E^{T}$, where $j\in n(i)$.
    Set $i$'s state to 1, which means it has been sampled. Then
    $\forall j\in n(i)$, if $j$ is not sampled, add it to the
    sampling queue $Q$, i.e., $Q=Q\cup\{j\}$. If $|E^{P}|/|E| \geq
    s_f$, then go to Step 4.
\item
    Step 3: $\forall q\in Q$, perform Step 2 with $i$ replaced by
    $q$.
\item
    Step 4: let $E^{P}=E-E^{T}$ and exit.
\end{itemize}

\emph{Metropolis-Hastings Random Walk(MHRW)} The Metropolis-Hastings
algorithm is a general Markov Chain Monte Carlo technique for
sampling from a probability distribution $\psi$ that is difficult to
sample from directly~\cite{mhrw_facebook}. It appropriately modify
the transition probabilities so that it converges to the desired
uniform distribution~\cite{mhrw_facebook}. The procedure of the
method could be depicted as
\begin{itemize}
\item
    Step 1: randomly select a starting node $i$ from $V$.
\item
    Step 2: randomly select the next hop $j$ from $n(i)$. Let
    $p_{ij}=k_i/k_j$. Generate a uniform $p\in[0,1]$, if $p\leq
    \min\{1.0, p_{ij}\}$, add $e_{ij}\in E$ to $E^T$, else just let $j=i$. If
    $|E^T|/|E|\geq s_f$, then go to Step 4.
\item
    Step 3: repeat Step 2 with $i$ replaced by $j$.
\item
    Step 4: let $E^{P}=E-E^{T}$ and exit.
\end{itemize}
It can be easily obtained in Step 2 that in MHRW, it tries to avoid
the situation of biased to the nodes with higher degrees.

\emph{Frontier Sampling(FS)} FS is proposed to implement
multidimensional random walks in the networks~\cite{fs}. It performs
$m$ dependent random walks in the network and $m$ is denoted as its
dimension. The procedure of this method could be presented as
\begin{itemize}
\item
    Step 1: randomly select $m$ nodes from $V$ and add them to the
    seed list $S$.
\item
    Step 2: select a node $i$ from $S$ with the probability
    $p_i=k_i/\sum_{q\in S}{k_q}$. Then randomly select $j$ from
    $n(i)$, add $e_{ij}\in E$ to $E^T$ and replace $i$ by $j$ in
    $S$. If $|E^{T}|/|E|\geq s_f$, go to Step 3, else repeat this
    step.
\item
    Step 3: let $E^{P}=E-E^{T}$ and exit.
\end{itemize}

\emph{Forest Fire(FF)} FF is proposed in~\cite{ff} for sampling
large-scale networks to empirically analyze their static or dynamic
graph properties. This method can also used to generated networks as
an evolution model~\cite{ff_graphovertime}. Its implementation could
be depicted briefly as
\begin{itemize}
\item
    Step 1: randomly select a seed node from $V$ and add it to the
    \emph{burnlist} $B$.
\item
    Step 2: get a node $i$ from $B$, add  all the edges connected to $i$ to $E^{T}$ and set it as the burned node. Then generate a
    random number $\beta$ that is geometrically distributed with
    mean
    \begin{equation}
    \frac{p_f}{1-p_f},
    \label{eq:mean_pf}
    \end{equation}
    where $p_f$ is the forward-burning probability. Select $\min\{\beta, k_i\}$ neighbors from
    $n(i)$ that are not yet burned and add them to $B$.
    If$|E^{T}|/|E|\geq s_f$, go to Step 3, else repeat this step.
\item
    Step 3: let $E^{P}=E-E^{T}$ and exit.
\end{itemize}

\emph{Pure Random(PR)} In order to compare with the conventional
random selections of the training set and the probe set, here we
import an ideal pure random sampling method, which is generally not
feasible in the practical scenario. We assume that the global view
of the network is obtained then we would randomly select $s_f$ links
and add them to $E^T$, while the remaining ones belong to $E^P$.

In summary, we would employ these five sampling methods to evaluate
the ten utilized prediction measures on the real-world networks,
which are going to be introduced in the next section.

\section{Real-world Data Sets}
\label{sec:datasets}

In this section, nine real-world data sets we utilize in the present
work would be depicted in detail. It is worthy to be noted that for
all the data sets, for the reason of sampling, we only perform
experiments on their giant connected components.

These networks come from different fields in the real word.
\texttt{Netscience} is a network of co-authorships between
scientists who are themselves publishing on the topic of network
science~\cite{netscience}. texttt{Power} is a well-connected
electrical power grid of western US, where nodes denote generators,
transformers and substations and edges denote the transmission lines
between them~\cite{Grid}. \texttt{USAir} is the network of US air
transportation system, in which the nodes are airports while the
links are the airlines among
them\footnote{http://vlado.fmf.uni-lj.si/pub/networks/data/mix/USAir97.net}.
\texttt{Yeast} is an network of protein-protein
interaction\footnote{http://vlado.fmf.uni-lj.si/pub/networks/data/bio/Yeast/Yeast.htm}.
\texttt{Dimes} is a topology of the Internet int the level of
Autonomous System(AS) and comes from the project of
DIMES\footnote{http://www.netdimes.org}. The AS-level data set we
use was released at March, 2010. In this network, each node
represents an AS, while each link means there exists an AS path
between the related two nodes. \texttt{Pb} is a directed network of
US political
blogs\footnote{http://www-personal.umich.edu/~mejn/netdata/polblogs.zip}.
Here we treat its links as undirected and self-connections are
omitted. \texttt{Caltech} is the Facebook network whose ties are
within California Institute of Technology, in which a node is a user
and the friendships between users are the links~\cite{caltech}.
\texttt{Email} covers all the email communication within a data set
of around half million emails. Nodes of the network are email
addresses and if an address $i$ sent at least one email to address
$j$, the graph contains an undirected edge from $i$ to
$j$~\cite{email}. \texttt{Hepph} is from the e-print
arXiv\footnote{http://www.arxiv.org} and covers scientific
collaborations between authors with papers submitted to High Energy
Physics-Phenomenology category from January 1993 to April 2003. If
an author $i$ co-authored a paper with author $j$, the graph
contains a undirected edge from $i$ to $j$~\cite{ff_graphovertime}.

The basic topological characteristics of these data sets are listed
in Table~\ref{tab:dataset}.

\begin{table*} [th]
\centering \caption{Real-world Data Sets.}
\begin{tabular}{|l|l|l|l|l|l|}
\hline Data Set & $\arrowvert V\arrowvert$ & $\arrowvert
E\arrowvert$
& $\langle k\rangle$ & $C$ & $H$\\
\hline \hline
\texttt{Netscience} & 379 & 914 & 4.82 & 0.74 & 1.66\\
\texttt{Power} & 4 941 & 6 594 & 2.67 & 0.08 & 1.45\\
\texttt{USAir} & 332 & 2126 & 12.81 & 0.63 & 3.46\\
\texttt{Yeast} & 2 224 & 6 609 & 5.94 & 0.14 & 2.80\\
\texttt{Dimes} & 26 424 & 90 267 & 6.83 & 0.47 & 74.66\\
\texttt{Pb} & 1 222 & 16 714 & 27.36 & 0.32 & 2.97\\
\texttt{Caltech} & 762 & 16 651 & 43.70 & 0.41 & 1.72\\
\texttt{Email} & 33 696 & 180 811 & 10.73 & 0.50 & 13.27\\
\texttt{Hepph} & 11 204 & 117 619 & 21.00 & 0.62 & 6.23\\
\hline
\end{tabular}
\label{tab:dataset}
\end{table*}

\section{Evaluation from the view of sampling}
\label{sec:results}

In the section, we perform evaluating experiments on different
real-world data sets and unveil that for different sampling methods,
each prediction measure performs unevenly. Finally, we also discuss
the evaluation of performance in the situation of tuning sampling
parameters.

\subsection{Typical Evaluation}
\label{subsec:evaluation}

For each data set, we sample 100 times and generate 100 partitions
of $E^T$ and $E^P$ with $s_f=0.9$. Then for the 100 cases, we employ
each prediction measure to uncover the missing links and get the
averaged AUC as its performance. For FS and FF, we set $m=100$ and
$p_f=0.8$ in the following experiments in this subsection. The other
configurations of these parameters would be discussed in the latter
one.

\begin{figure*}
\centering
 \subfloat[\texttt{Netscience}]{\includegraphics[width=1.8in]{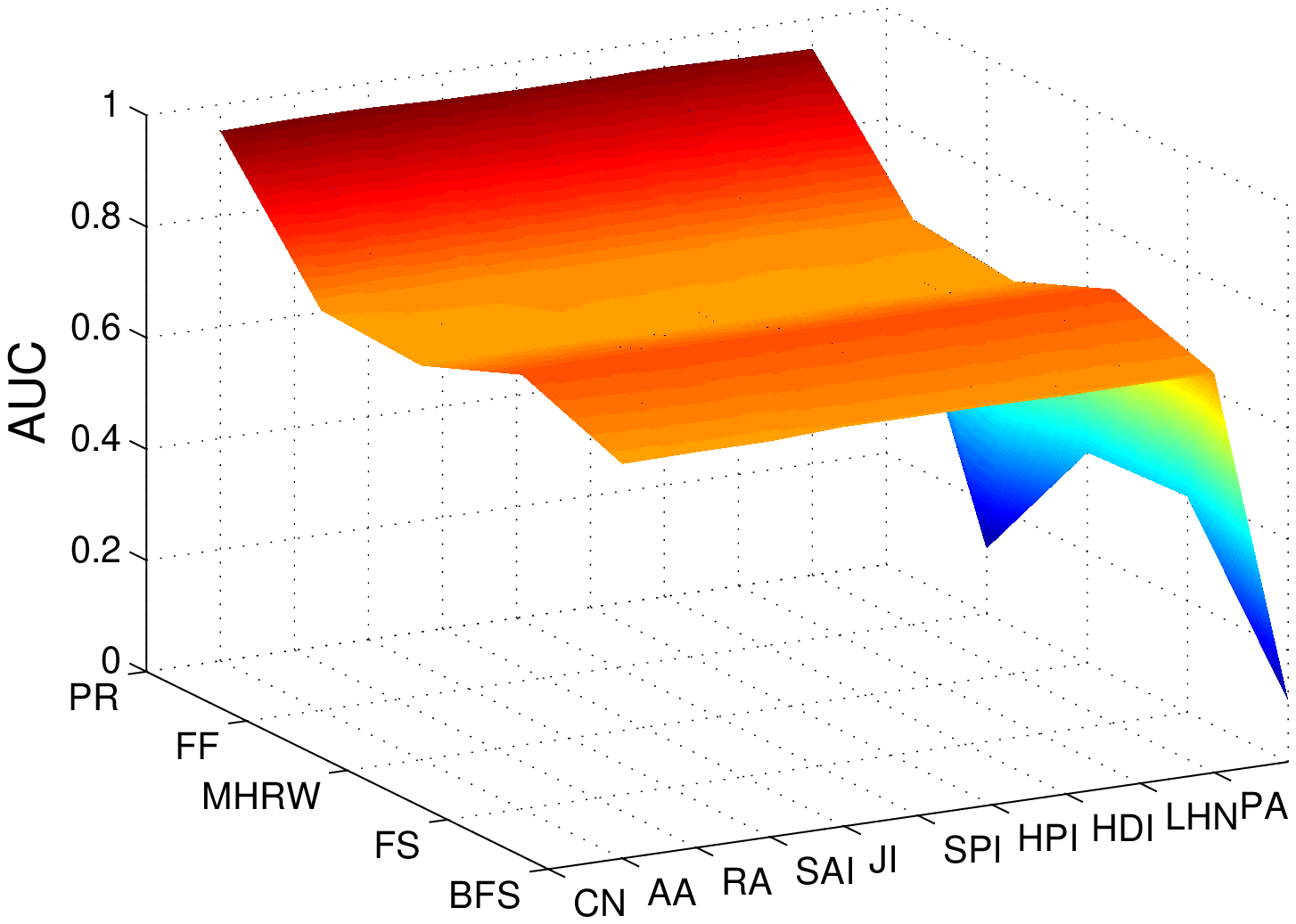}
 \label{fig:netscience_auc}}
 \subfloat[\texttt{Power}]{\includegraphics[width=1.8in]{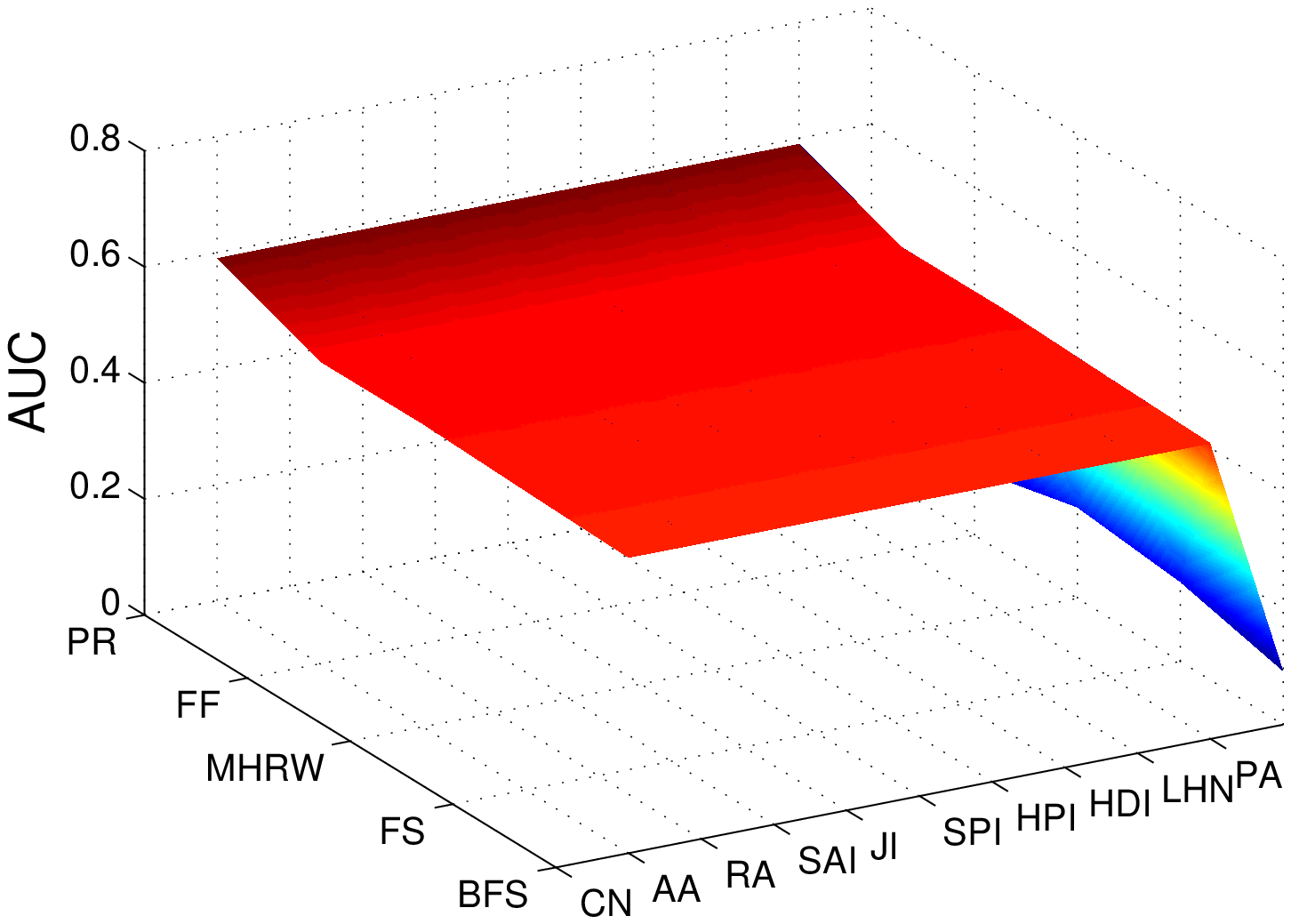}
 \label{fig:power_auc}}
 \subfloat[\texttt{USAir}]{\includegraphics[width=1.8in]{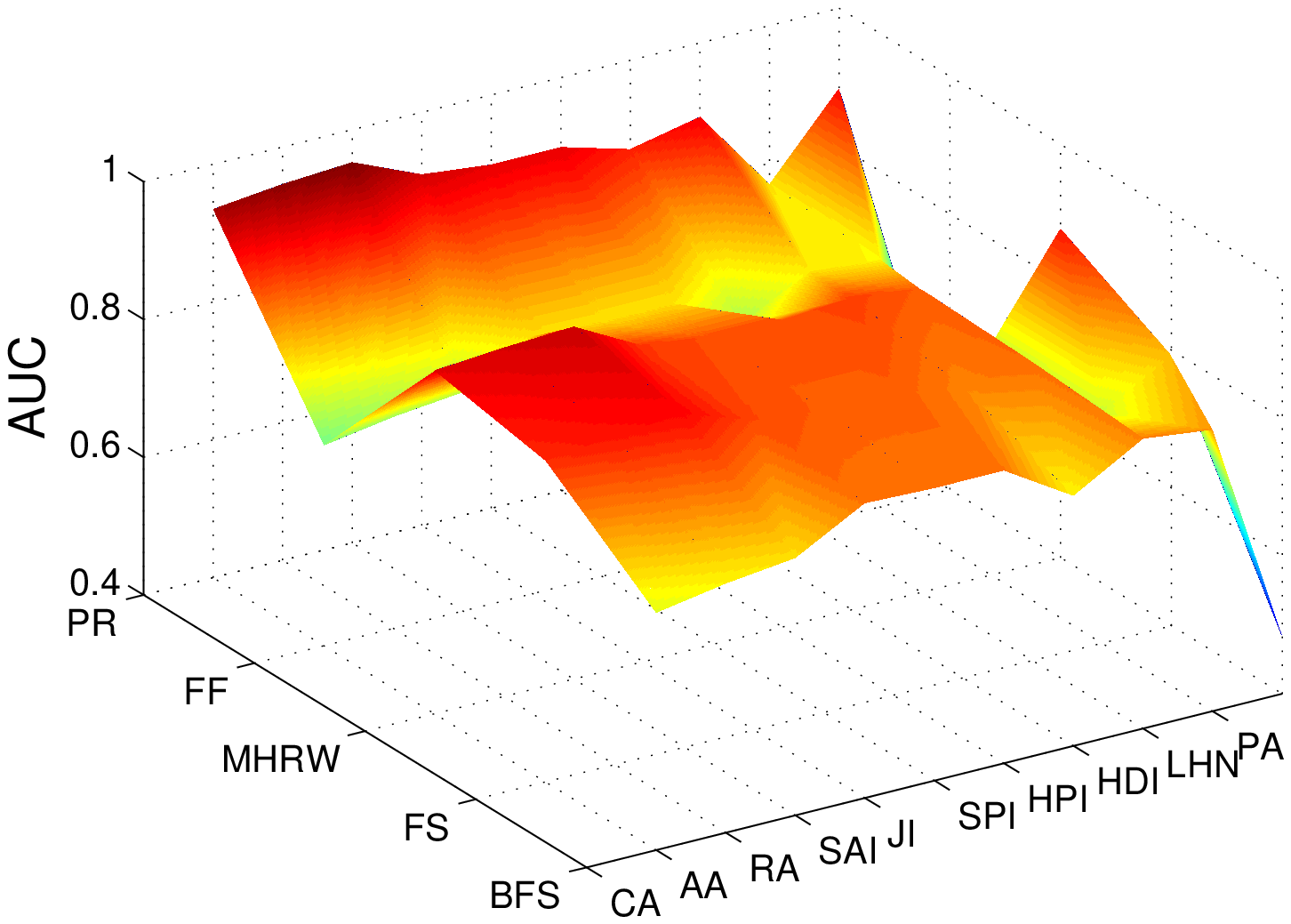}
 \label{fig:usair_auc}}\\
  \subfloat[\texttt{Yeast}]{\includegraphics[width=1.8in]{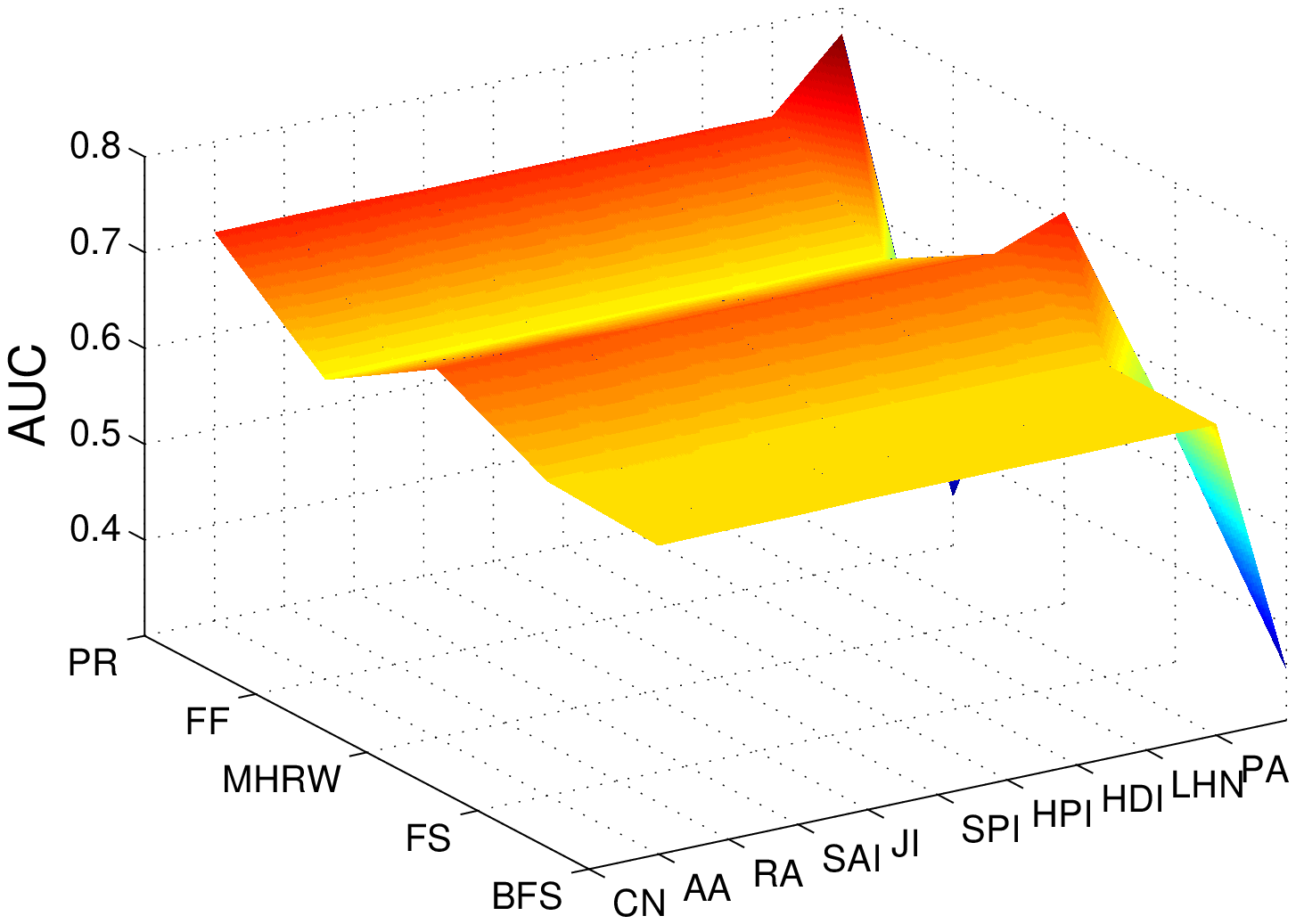}
 \label{fig:yeast_auc}}
  \subfloat[\texttt{Dimes}]{\includegraphics[width=1.8in]{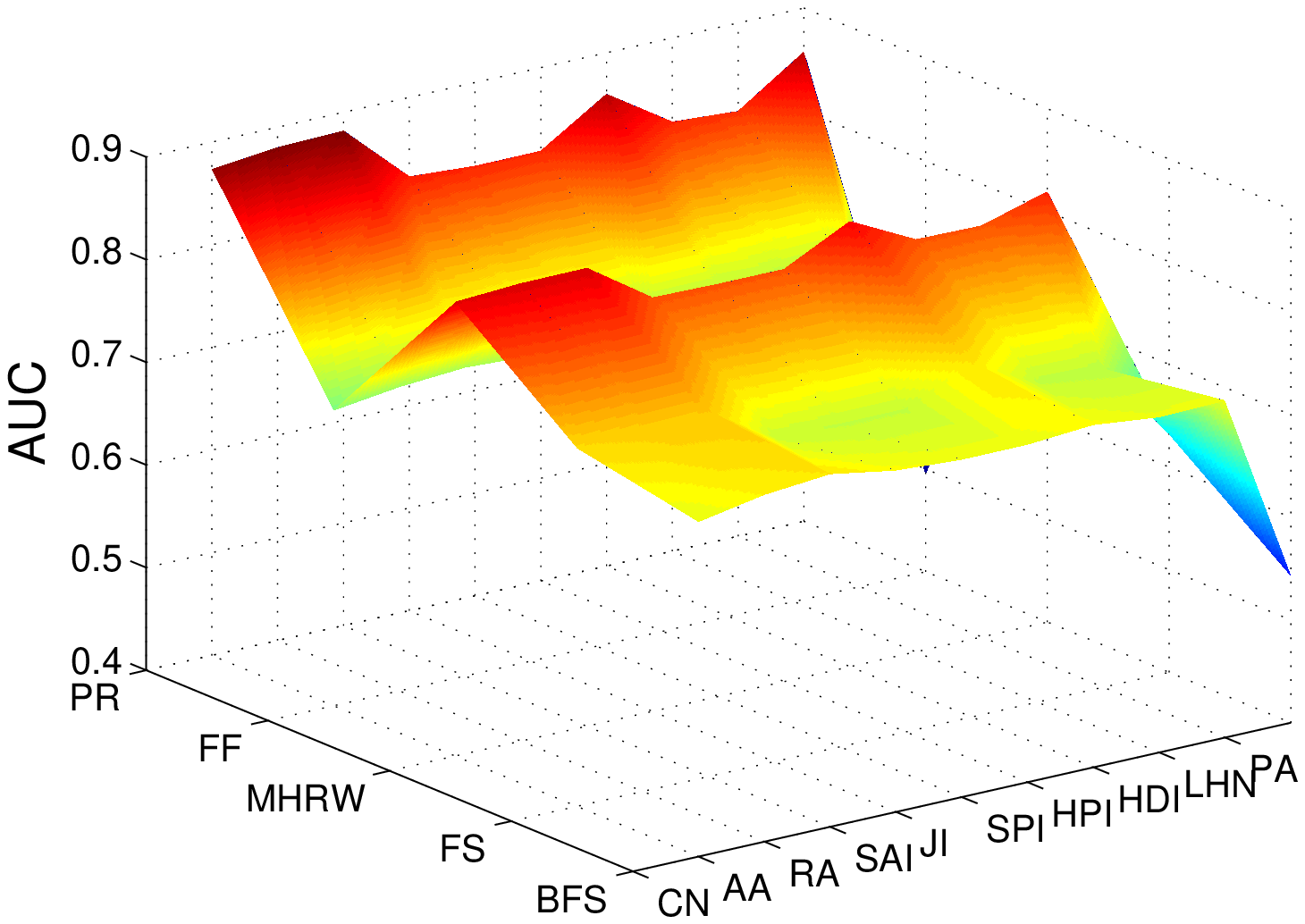}
 \label{fig:dimes_auc}}
  \subfloat[\texttt{Pb}]{\includegraphics[width=1.8in]{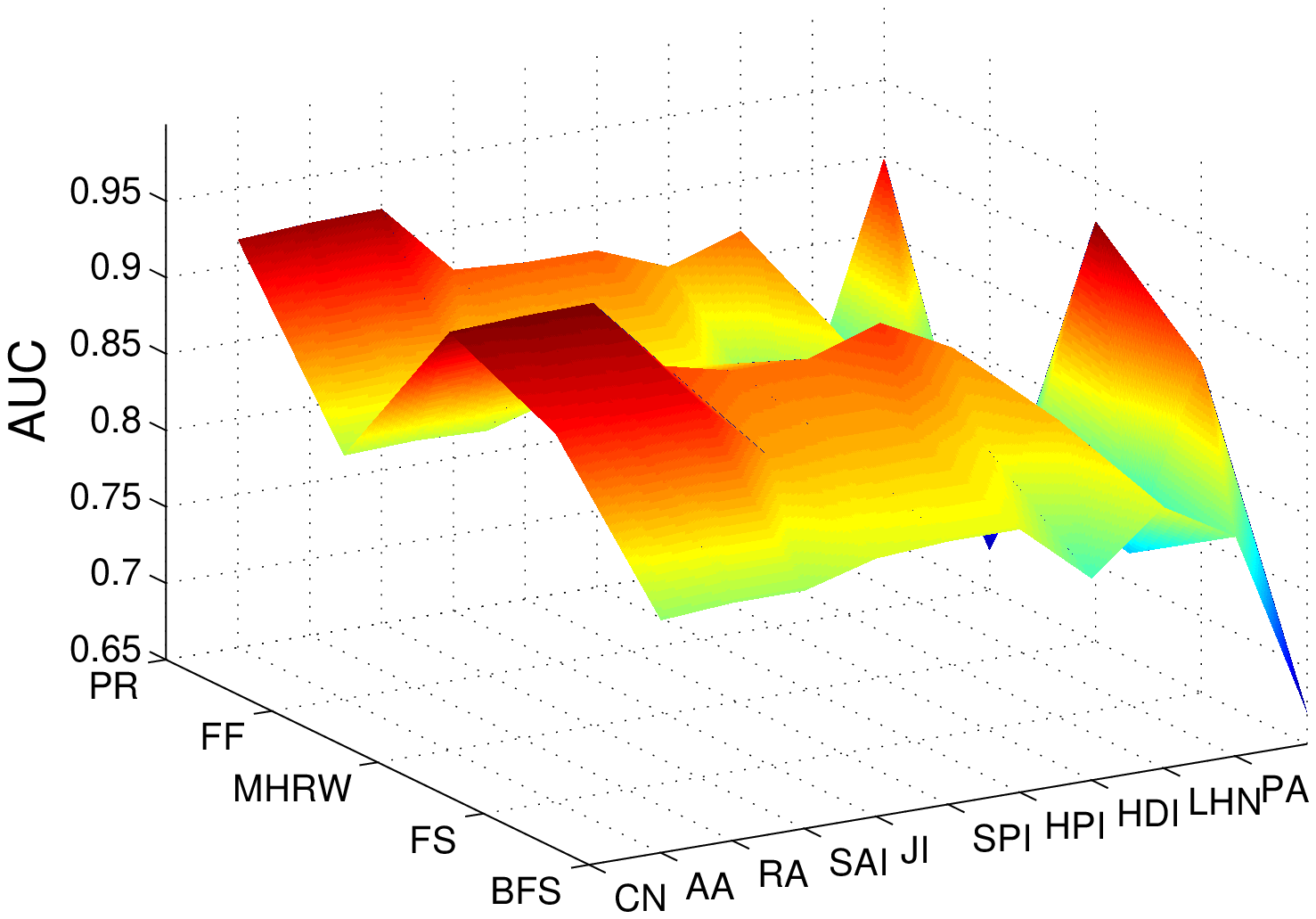}
 \label{fig:pb_auc}}\\
 \subfloat[\texttt{Caltech}]{\includegraphics[width=1.8in]{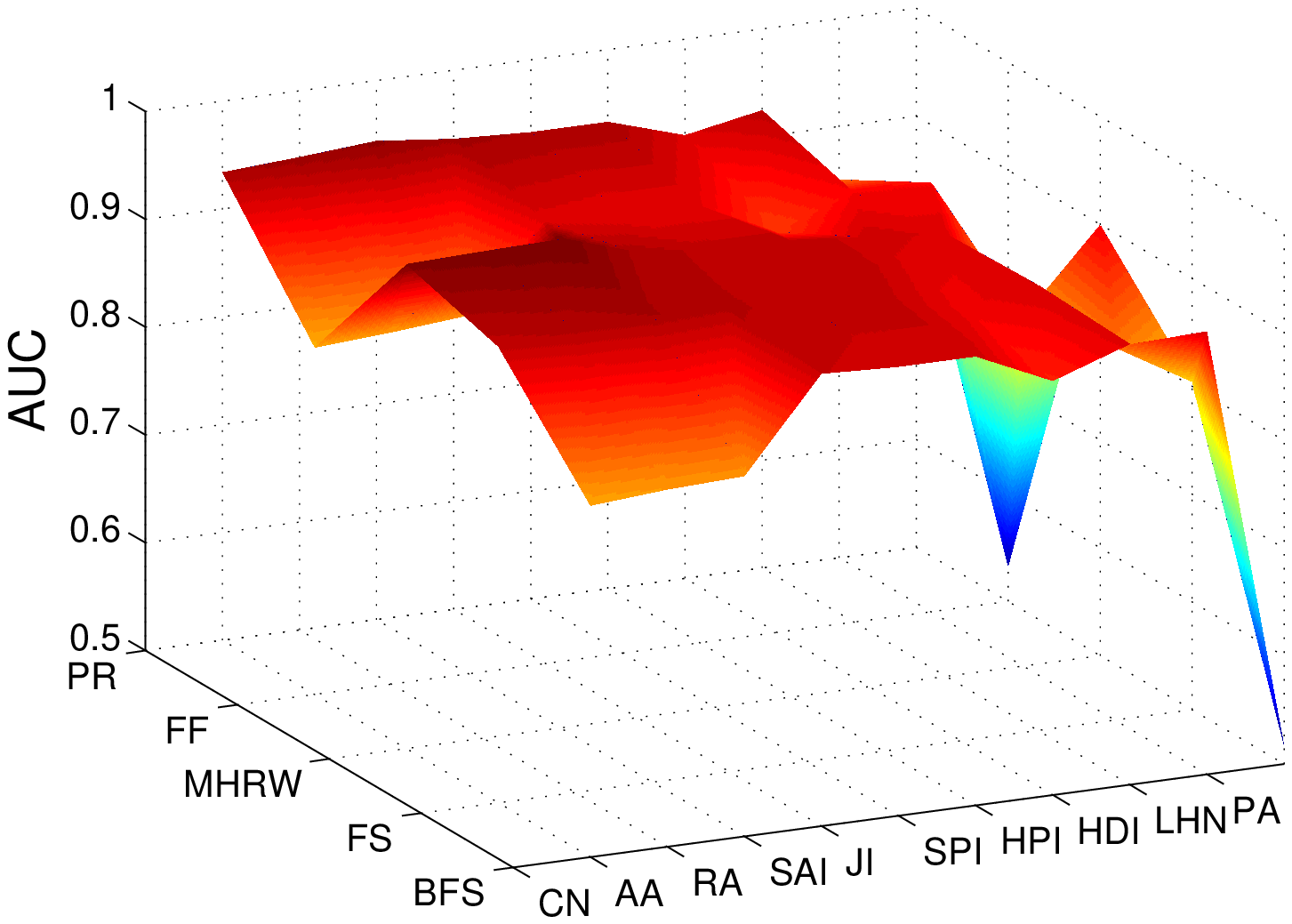}
 \label{fig:caltech_auc}}
  \subfloat[\texttt{Email}]{\includegraphics[width=1.8in]{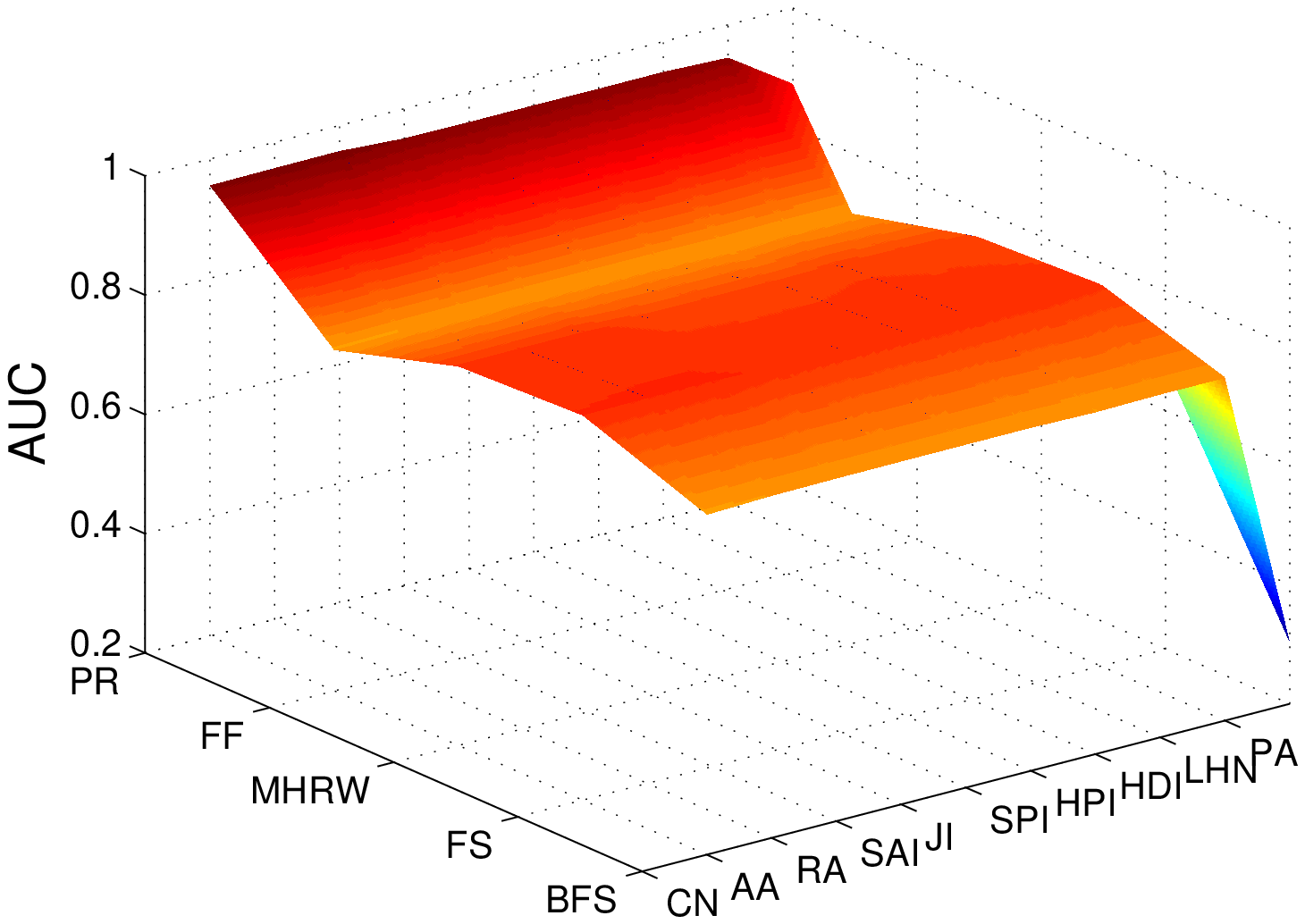}
 \label{fig:email_auc}}
  \subfloat[\texttt{Hepph}]{\includegraphics[width=1.8in]{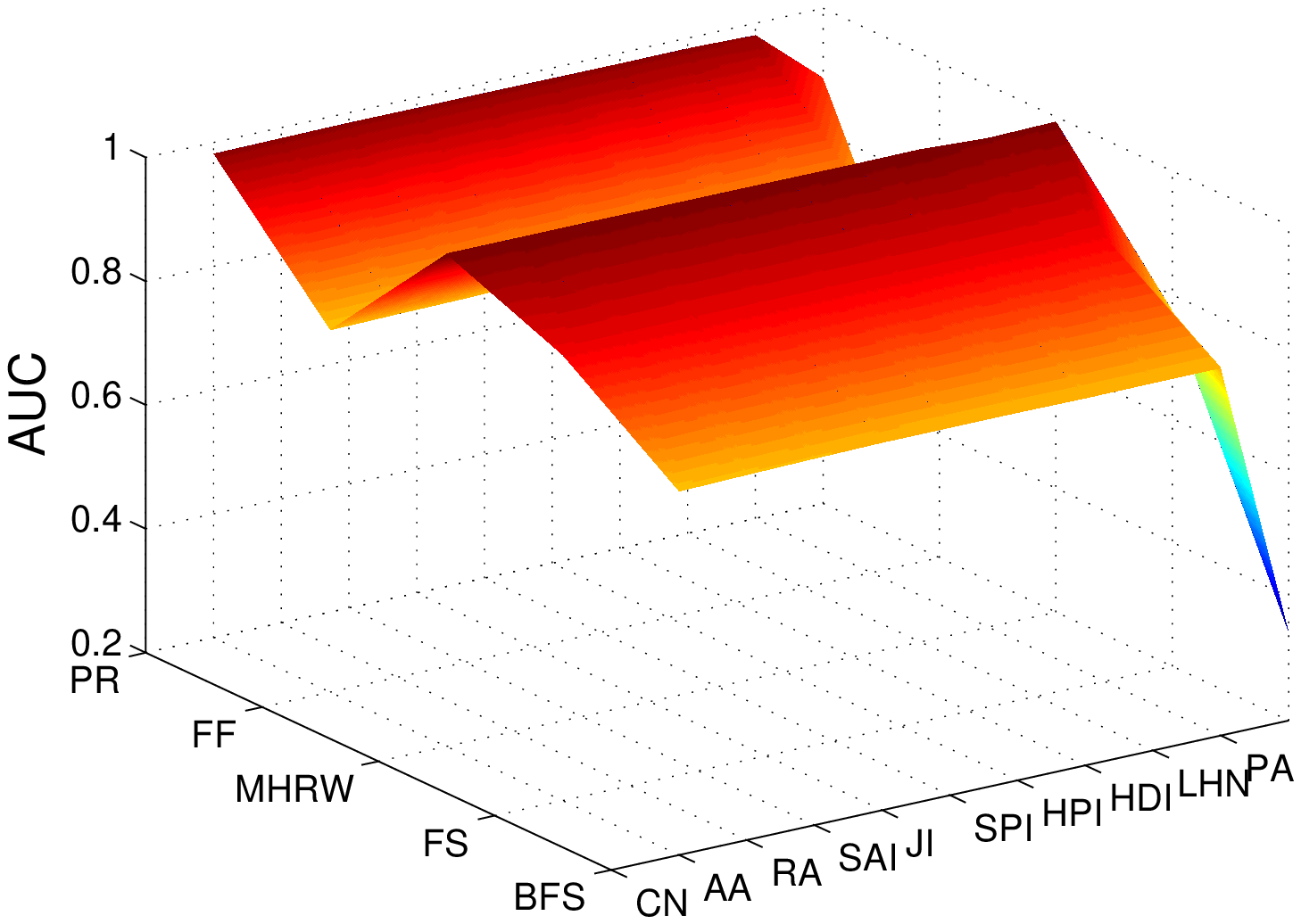}
 \label{fig:hepph_auc}}\\
\caption{Different prediction measures perform on $E^T$ generated by
different sampling methods.} \label{fig:evaluation_auc}
\end{figure*}

As shown in Figure~\ref{fig:evaluation_auc}, each prediction method
performs unevenly for different sampling approaches, particular for
PA. For an instance, as shown in Figure~\ref{fig:pb_auc}, AUC of CN
on \texttt{Dimes} is 0.73 when the probe set is obtained by BFS,
while it turns to be 0.75, 0.84, 0.69 and 0.87 as the probe set is
determined by FS, MHRW, FF and PR, respectively. With respect to PA,
it performs best for PR with AUC equals to 0.86, however, for other
sampling methods, its AUC decreases, e.g., 0.54, 0.63, o.82 and 0.50
for BFS, FS, MHRW and FF,respectively. It is also indicated in
Figure~\ref{fig:evaluation_auc} that for most of the data sets we
employ in this work, all the prediction methods perform best when
the probe set is obtained through the conventional sampling way,
i.e., PR. This tells us that in the previous work, the performance
of local information based link predictions might be overestimated.

\begin{figure*}
\centering
 \subfloat[\texttt{Caltech}]{\includegraphics[width=1.8in]{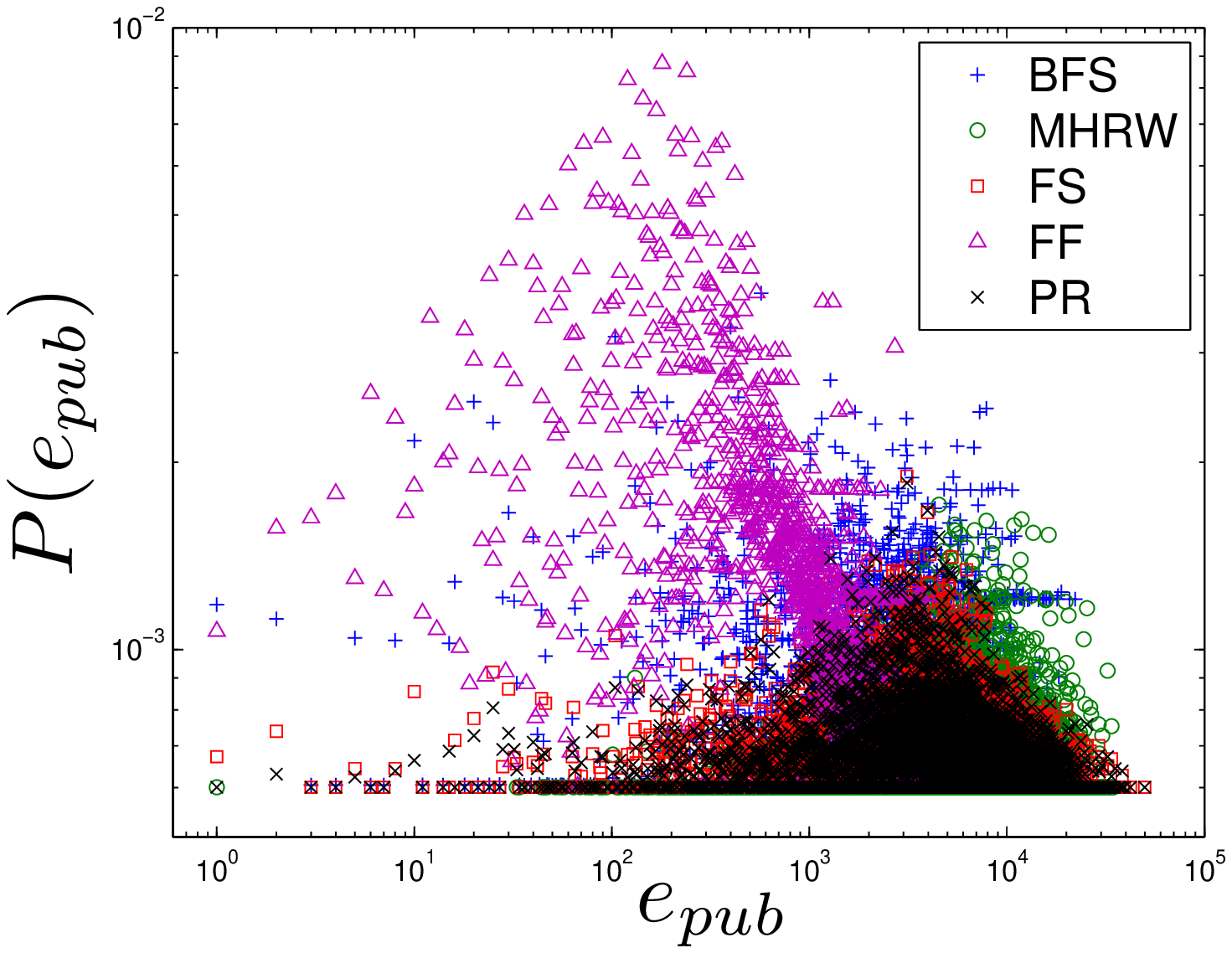}
 \label{fig:caltech_epub}}
 \subfloat[\texttt{Email}]{\includegraphics[width=1.8in]{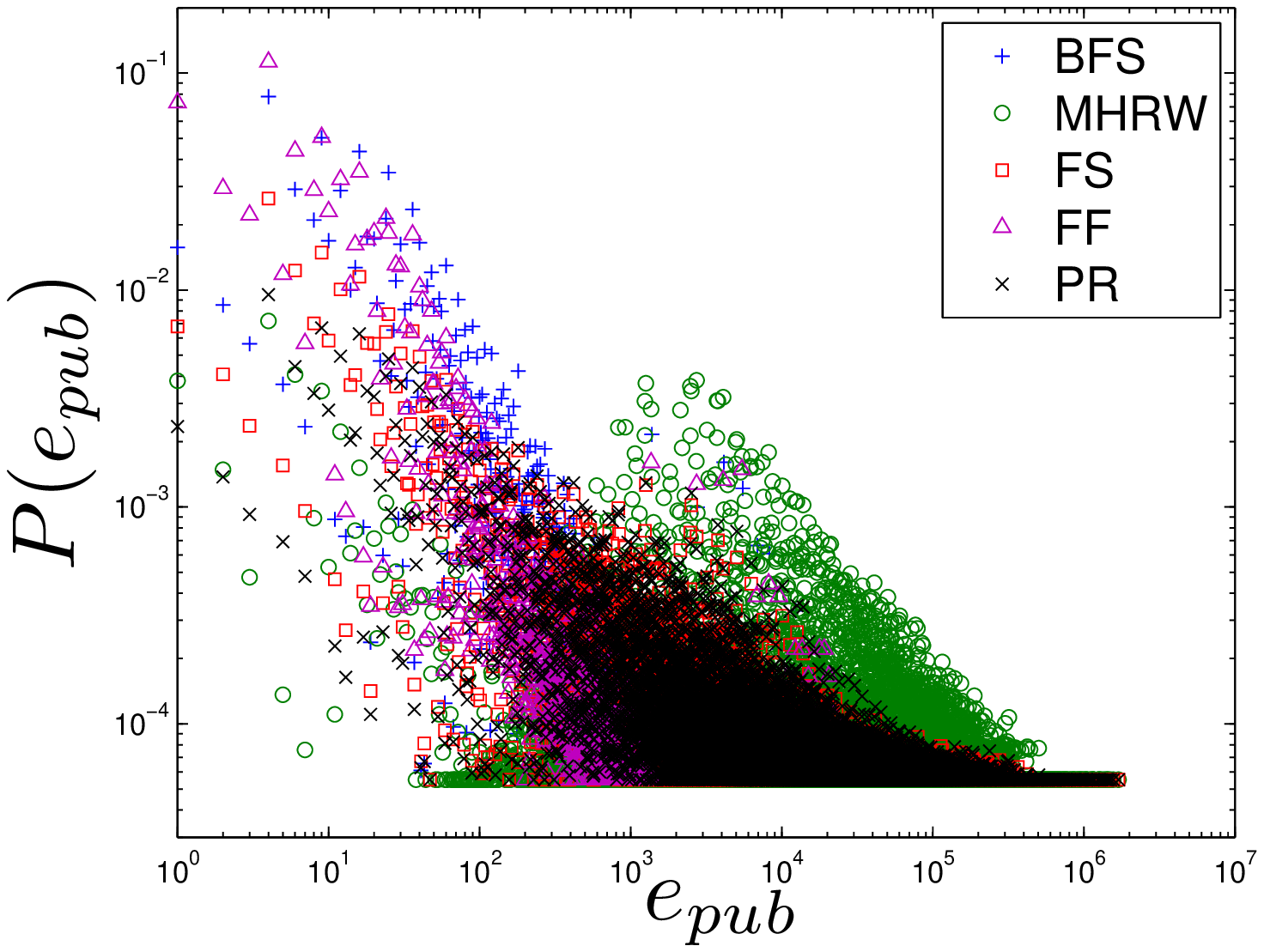}
 \label{fig:email_epub}}
 \subfloat[\texttt{USAir}]{\includegraphics[width=1.8in]{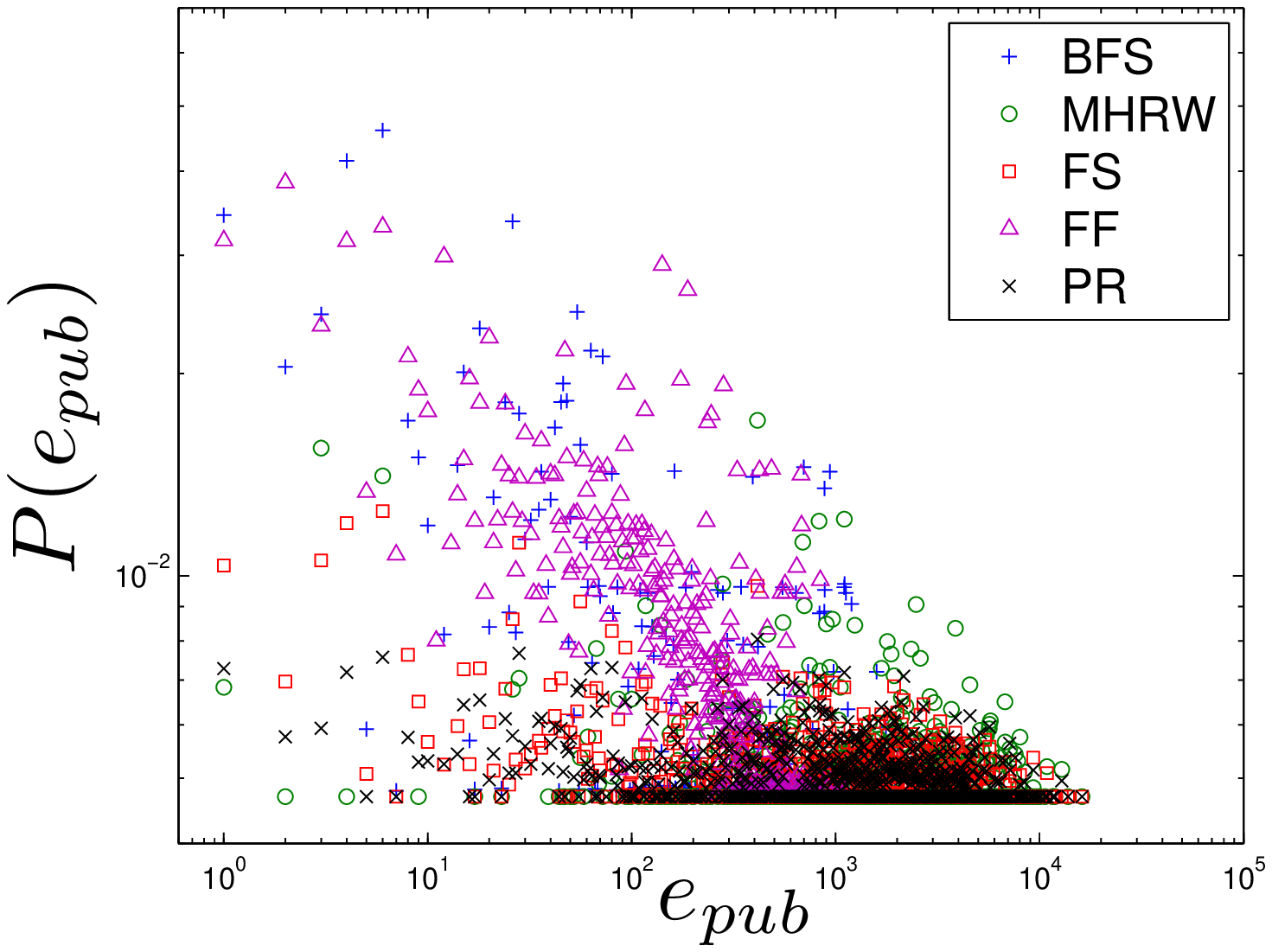}
 \label{fig:usair_epub}}\\
 \caption{The distribution of $e_{pub}(i,j)$ in the probe set. For each randomly selected data set,
 we obtain 100 probe sets through each sampling methods and get the averaged distribution.}
 \label{fig:epub_distribution}
\end{figure*}

\begin{figure*}
\centering
  \subfloat[\texttt{Yeast}]{\includegraphics[width=1.8in]{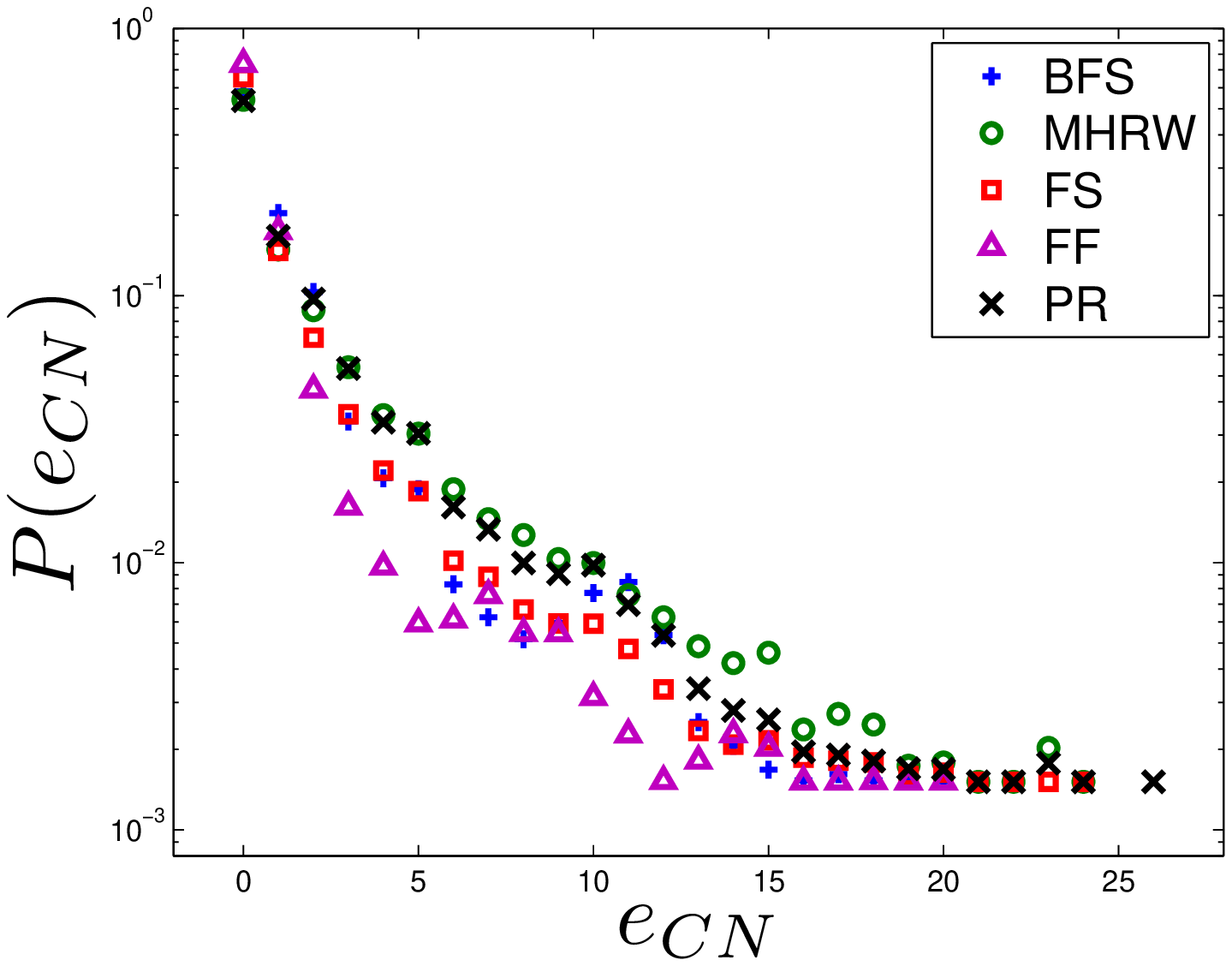}
 \label{fig:yeast_ecn}}
  \subfloat[\texttt{Dimes}]{\includegraphics[width=1.8in]{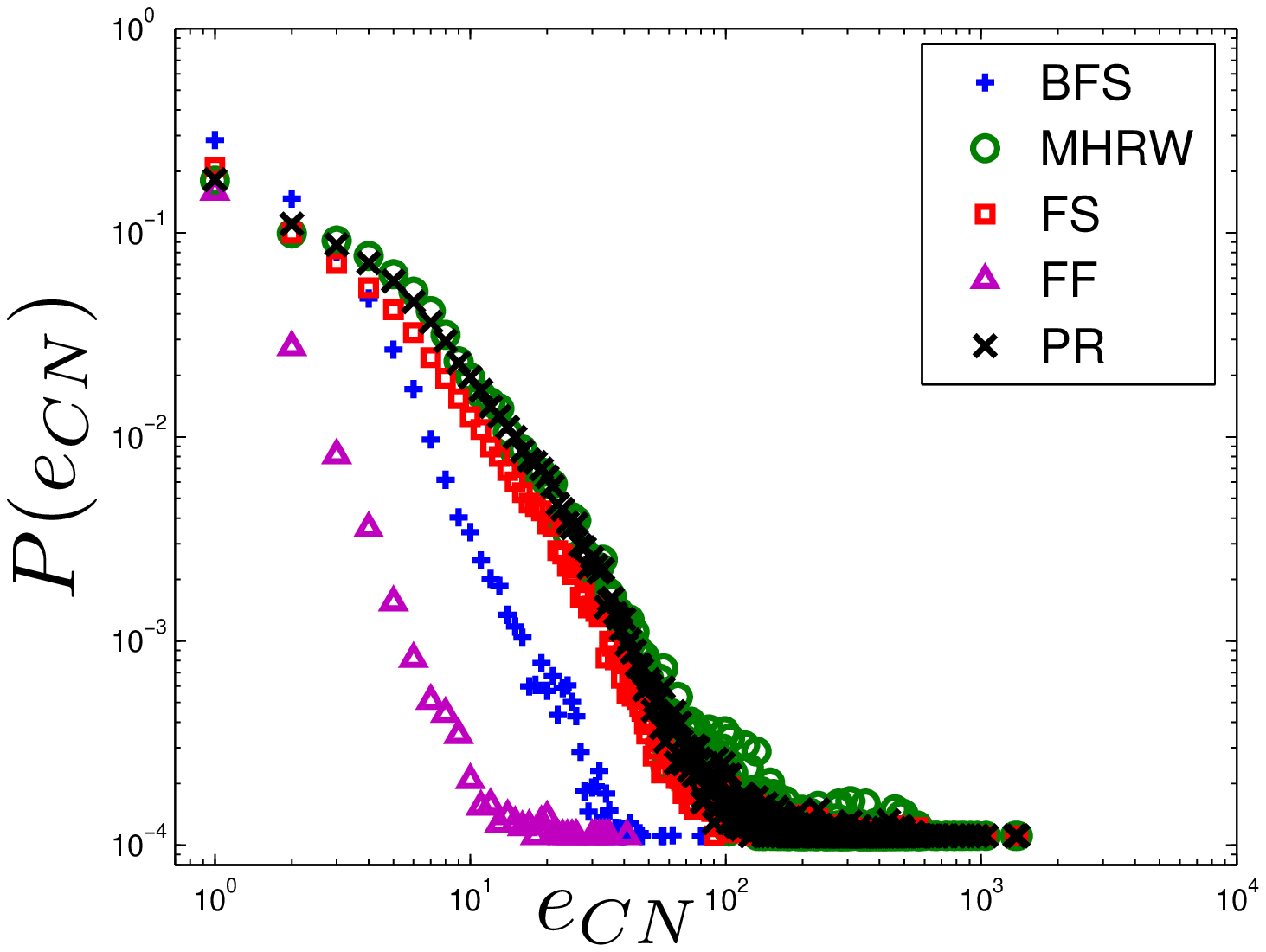}
 \label{fig:dimes_ecn}}
  \subfloat[\texttt{Pb}]{\includegraphics[width=1.8in]{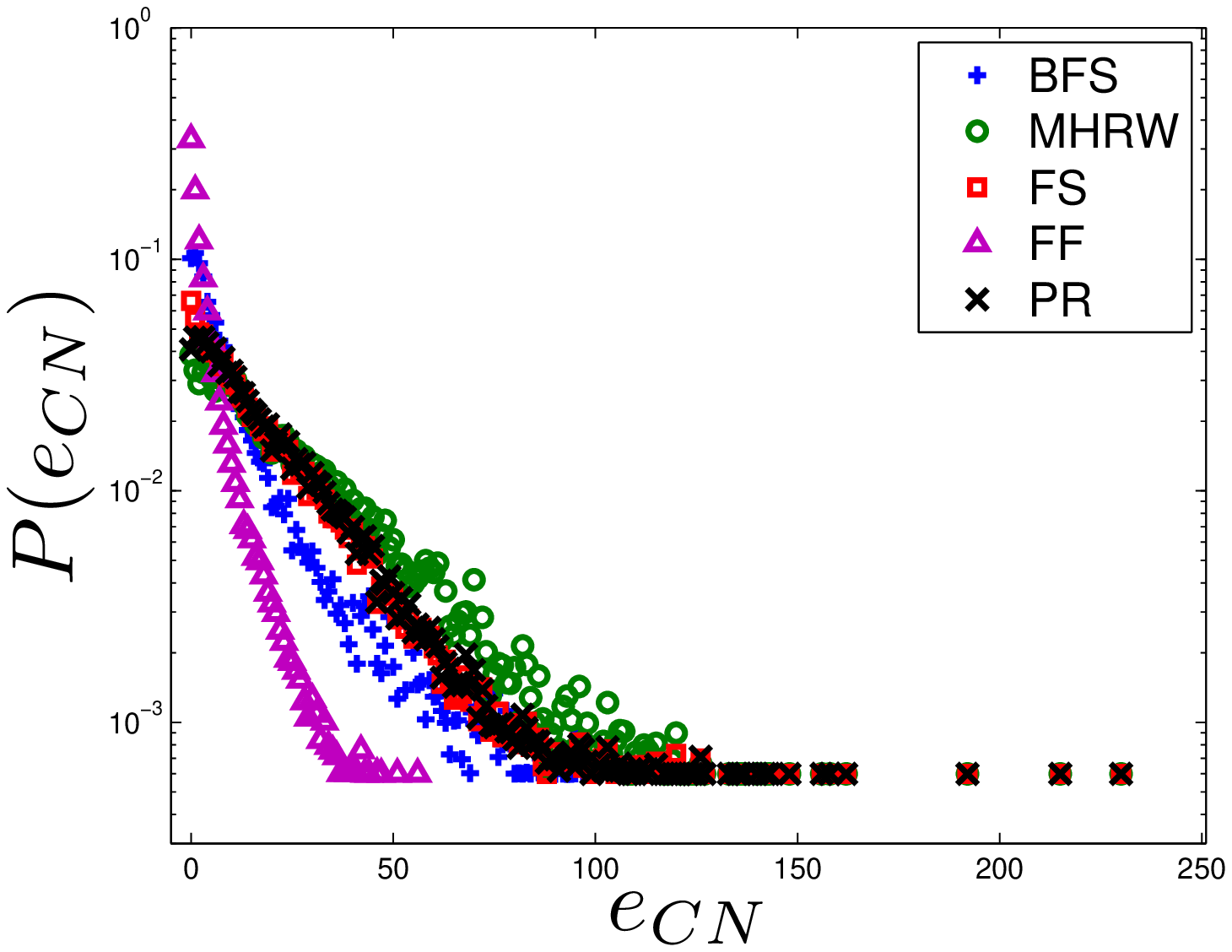}
 \label{fig:pb_ecn}}\\
\caption{The distribution of $e_{CN}(i,j)$ in the probe set. For
each randomly selected data set, we obtain 100 probe sets through
each sampling methods and get the averaged distribution.}
\label{fig:ecn_distribution}
\end{figure*}

As mentioned in~\ref{subsec:linkprediction}, most of these measures
are related with $e_{pub}(i,j)$ and $e_{CN}(i,j)$. In order to
illustrate the diversity of the performance for different sampling
methods, we observe the distribution of $e_{pub}(i,j)$ and
$e_{CN}(i,j)$, denoted as $P(e_{pub})$ and $P(e_{CN})$,
respectively, in the probe set. As can be seen in
Figures~\ref{fig:epub_distribution} and~\ref{fig:ecn_distribution},
for each randomly selected data set, $P(e_{pub})$ and $P(e_{CN})$
fluctuate diversely for each sampling method. Generally, for PR, FS
and MHRW, most of edges that are not sampled occupy lower values of
$e_{pub}(i,j)$ and $e_{CN}(i,j)$, while for BFS and FF, the fraction
of edges with large $e_{pub}(i,j)$ and $e_{CN}(i,j)$ is less.
Because of this, these prediction methods performs better on the
probe set generated by PR, FS and MHRW, however, it is
correspondingly hard for them to uncover the links with lower
$e_{pub}(i,j)$ or $e_{CN}(i,j)$ in probe sets obtained through BFS
and FF.

\begin{table*}
\centering \caption{Best prediction measures for each sampling
method.}
\begin{threeparttable}
\begin{tabular}{|l|l|l|l|l|l|}
\hline
Data Set & BFS& MHRW & FS & FF & PR\\
\hline \hline
\texttt{Netscience} & JI & RA   & RA & SAI & RA\\
\texttt{Power} & All\tnote{a} & HPI & CN/AA/RA/HPI & SAI/JI/SPI & RA\\
\texttt{USAir} & SAI & RA & RA & SAI & RA\\
\texttt{Yeast} & JI & AA & RA & JI/SAI & RA\\
\texttt{Dimes} & RA & RA & RA & RA & RA\\
\texttt{Pb} & JI & RA & RA & JI/SPI & RA\\
\texttt{Caltech} & SAI & RA & RA &SAI & RA\\
\texttt{Email} & SAI & RA & RA & SAI & AA\\
\texttt{Hepph} & SAI & RA & RA & SAI & RA\\
\hline
\end{tabular}
\end{threeparttable}
\begin{tablenotes}
    \item[a] a. All measures perform similarly except for PA.
\end{tablenotes}
\label{tab:bset_p_s}
\end{table*}

From the above experiments we could also disclose the proper
prediction method for different sampling approaches. In fact, as
shown in Table~\ref{tab:bset_p_s}, for each sampling method, there
exist several best prediction methods. As can be seen, for PR, MHRW
and FS, \textbf{RA} perform best on nearly all the data sets, which
is consist with the evaluation from randomly selected probe
sets~\cite{RA}. However, for BFS and FF, \textbf{SAI} performs more
outstandingly that other prediction measures. It is also interesting
that for the measure of PA, it performs poorly on all the data sets
and for all the sampling methods, especially for BFS and FS.

\subsection{Tuning sampling parameters}
\label{subsec:tuningparameters}

\begin{figure*}
\centering
 \subfloat[\texttt{USAir}]{\includegraphics[width=1.8in]{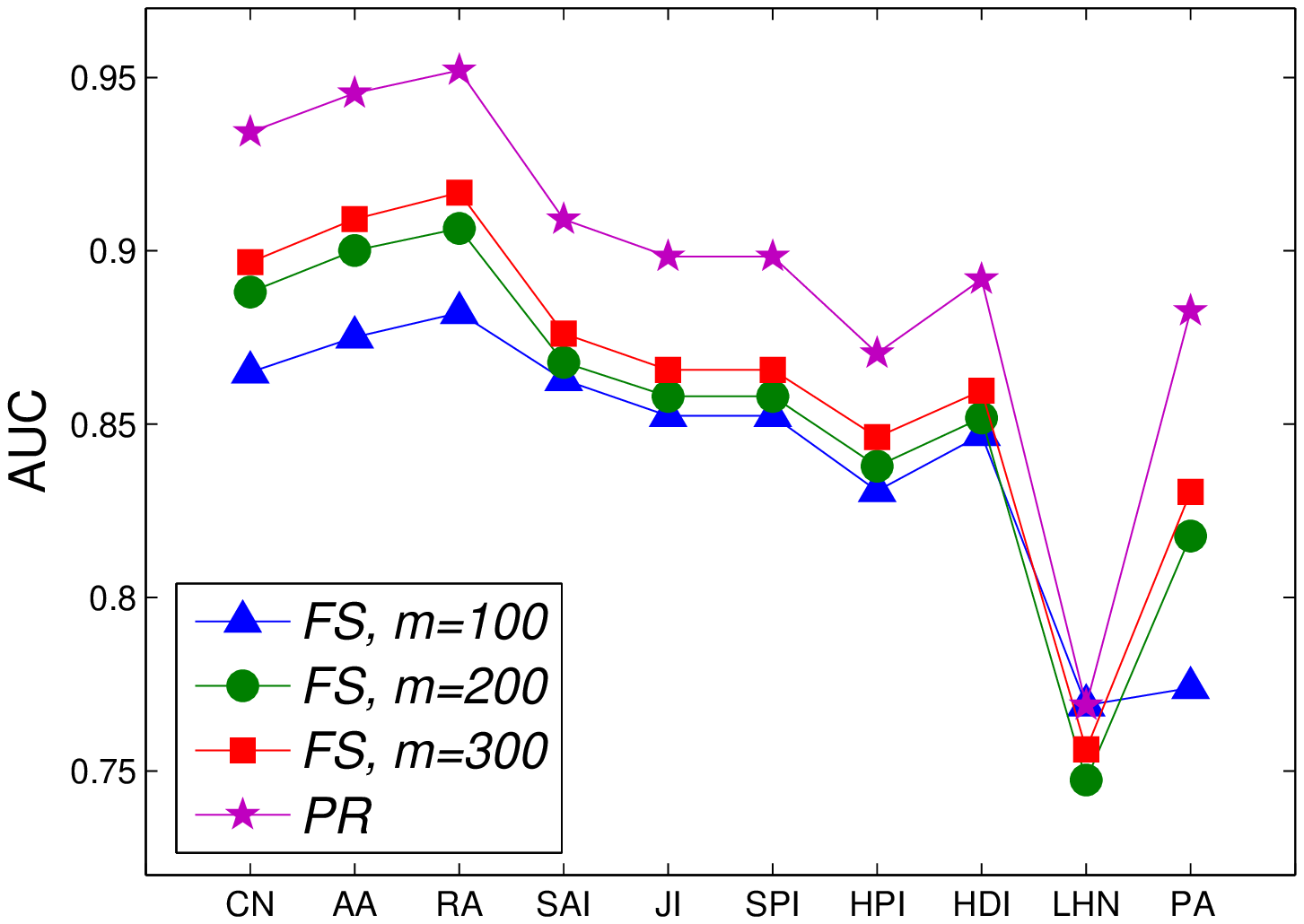}
 \label{fig:usair_fs_m}}
 \subfloat[\texttt{Dimes}]{\includegraphics[width=1.8in]{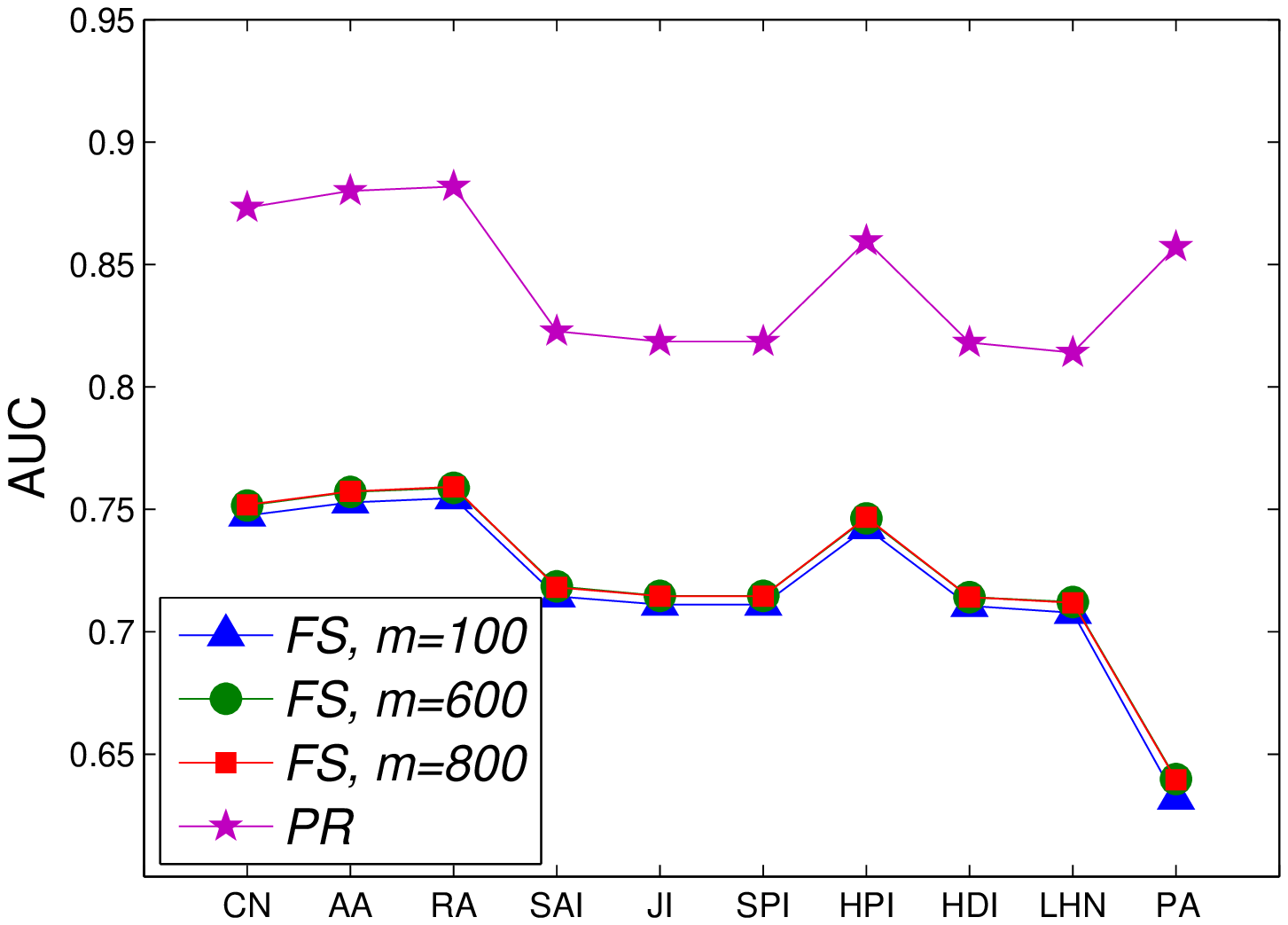}
 \label{fig:dimes_fs_m}}
 \subfloat[\texttt{Hepph}]{\includegraphics[width=1.8in]{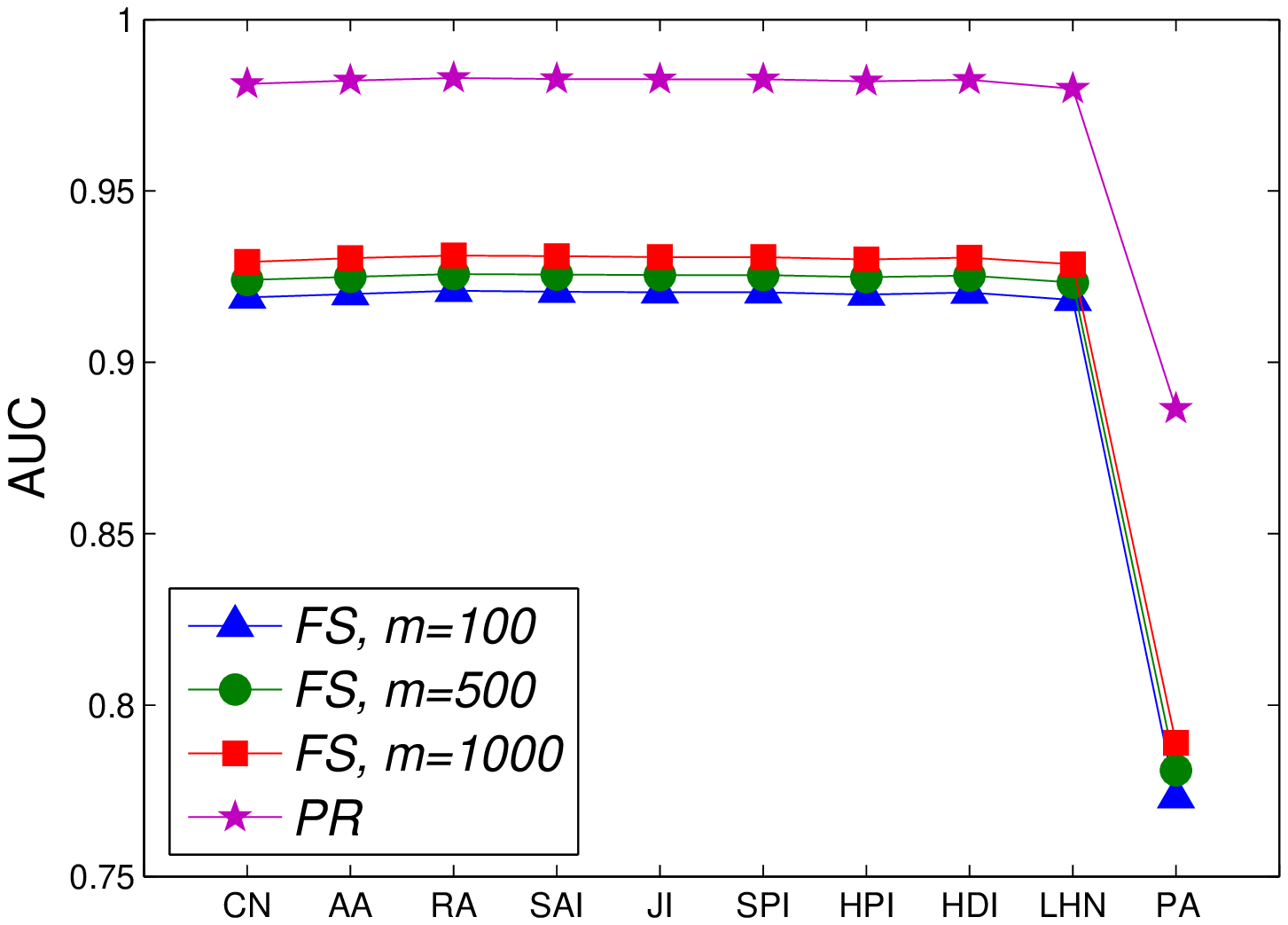}
 \label{fig:hepph_fs_m}}\\
 \caption{Evaluation of performance as $m$ varies for FS. For each $m$, we obtain 100 probe sets and get the averaged AUC
 as the final performance for each prediction measure.}
 \label{fig:fs_m}
\end{figure*}

\begin{figure*}
\centering
 \subfloat[\texttt{USAir}]{\includegraphics[width=1.8in]{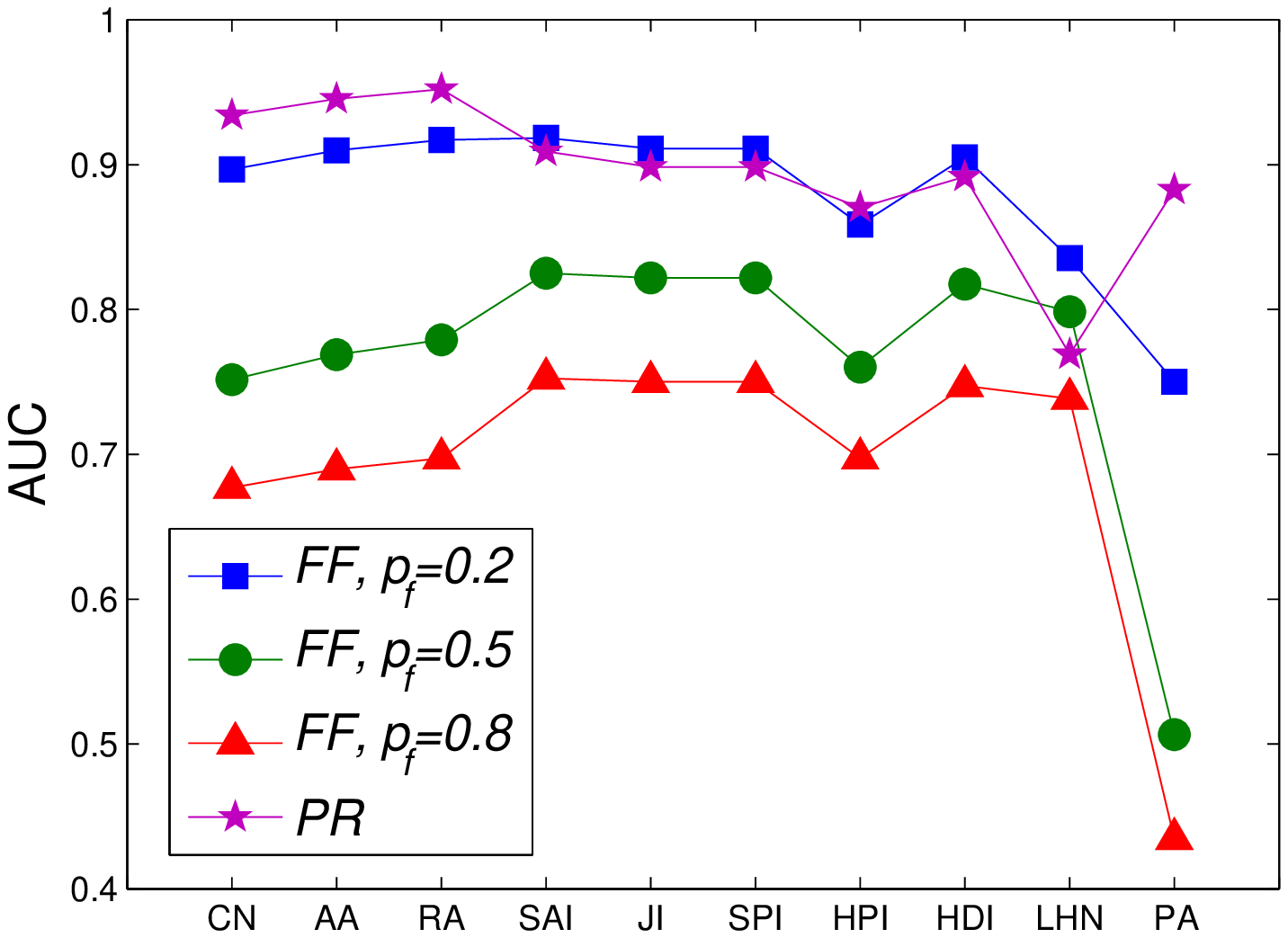}
 \label{fig:usair_ff_pf}}
 \subfloat[\texttt{Dimes}]{\includegraphics[width=1.8in]{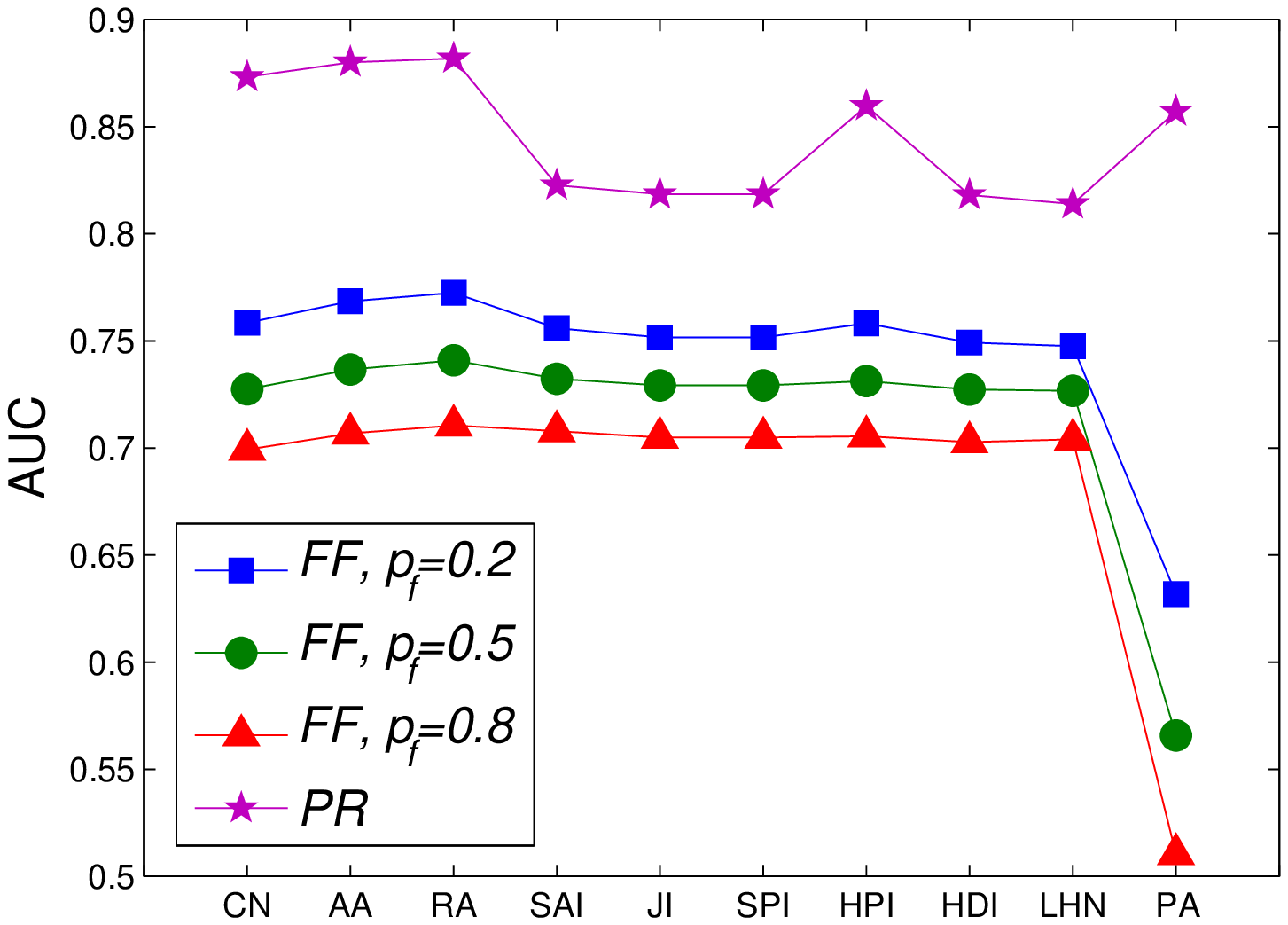}
 \label{fig:dimes_ff_pf}}
 \subfloat[\texttt{Hepph}]{\includegraphics[width=1.8in]{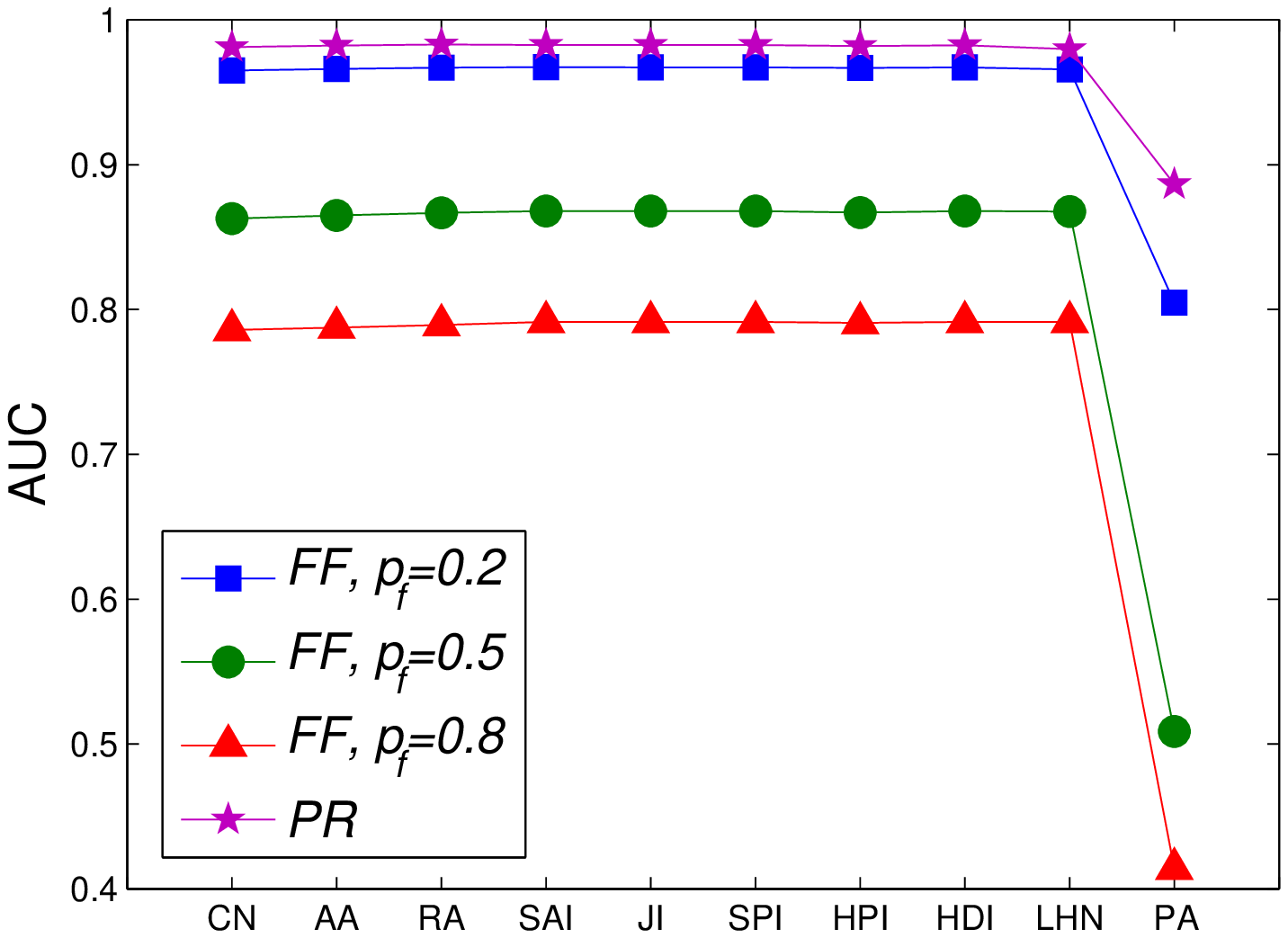}
 \label{fig:hepph_ff_pf}}\\
 \caption{Evaluation of performance as $p_f$ varies for FF. For each $p_f$, we obtain 100 probe sets and get the averaged AUC
 as the final performance for each prediction measure.}
 \label{fig:ff_pf}
\end{figure*}

In this subsection, we tune the sampling parameters for both FS and
FF with $s_f=0.90$ to observe variations of the performance.
Generally speaking, regarding to FS, large $m$ is favorable for
obtaining edges randomly from the network. However, in the real
world, it is hard to implement the random selection of large number
of seeds. Hence, in the following experiments, we only tune $m$ from
100 to $\min\{1000,|V|\}$.  With respect to FF, according to
Eq.~\ref{eq:mean_pf}, lower $p_f$ means at each sampling step, less
neighbors would be burned. In the following experiments,$p_f$ grows
from 0.2 to 0.8. We show the evaluation results of tuning $m$ and
$p_f$ from random selected three data sets in Figure~\ref{fig:fs_m}
and Figure~\ref{fig:ff_pf}, respectively. As shown in
Figure~\ref{fig:fs_m}, as $m$ grows, the performance of all the link
prediction measures increases and it also begin to saturate quickly
as $m$ goes up. However, as compared to PR, these measures still
perform poorly even for the case of $m=\min\{1000, |V|\}$. Similarly
for FF, as can be seen in Figure~\ref{fig:ff_pf}, when $p_f$
decreases from 0.8 to 0.2, AUC of all the prediction methods
increases gradually. Particularly, for the data set of
\texttt{USAir} as shown in Figure~\ref{fig:usair_ff_pf}, when
$p_f=0.2$, AUC of SAI, JI, SPI and LHN even exceeds the value
corresponding to the case of PR, however, the gap is little and
trivial. For other data sets we show here, the performance of these
measures is still weaker than the situation when the probe set is
determined by PR. It is also worthy to be noted that as $m$ or $p_f$
varies, the performance of PA fluctuates significantly and it tends
to perform better when the sampling method is PR.

In summary, as the sampling parameters varies, AUC obtained from FF
and FS increases closely to or a little bit higher than the value
from PR. It is still consist with our conjecture that PR supervised
division of the probe set and the training set would overestimate
the performance of prediction measures. Actually, large $m$ is
difficult to be satisfied, while smaller $p_f$ would make FF be much
time-consuming.

\section{Conclusion}
\label{sec:conclusion}

For the large-scale complex network in the real world, we could only
sample an incomplete picture of it, because of this, link prediction
methods have been employed to uncover the missing links in recent
years. However, in previous works, evaluations of these methods are
usually based on dividing the known edges randomly into two parts,
without considering that the pure random partition is even
impractical in the real world. For this reason, in this paper, we
try to reevaluate the performance of the local information based
link prediction measures reasonably from the view of several
sampling approaches that are pervasively utilized in reality. After
experiments on nine real-world data sets, we find that each of the
ten prediction measures performs unevenly for different sampling
methods. Particularly, for the conventional means, i.e., the pure
random sampling, these measures tend to perform best as compared
with other sampling approaches. It indicates that in the prior work,
the performance of the link prediction might be overestimated.
Finally, we conjecture that our findings could take a closer look at
the performance of the local information based prediction measures
and also shed light on the problem of how to select a proper
prediction method for a snapshot obtained through a certain sampling
approach.

\section*{Acknowledgment}
\label{sec:ack}

The research was supported by the fund of the State Key Laboratory
of Software Development Environment (SKLSDE-2011ZX-02).





\bibliographystyle{elsarticle-num}
\bibliography{refs}

\begin{thebibliography}{10}
\expandafter\ifx\csname url\endcsname\relax
  \def\url#1{\texttt{#1}}\fi
\expandafter\ifx\csname urlprefix\endcsname\relax\def\urlprefix{URL }\fi
\expandafter\ifx\csname href\endcsname\relax
  \def\href#1#2{#2} \def\path#1{#1}\fi

\bibitem{local_bayes}
Z.~Liu, Q.-M. Zhang, L.~Linyuan, T.~Zhou, Link prediction in complex networks:
  a local na\"{i}ve bayes model, arXiv:1105.4005v1.

\bibitem{mislove_sns_bfs}
A.~Mislove, M.~Marcon, K.~P. Gummadi, P.~Druschel, B.~Bhattacharjee,
  Measurement and analysis of online social networks, in: Proceedings of the
  7th ACM SIGCOMM conference on Internet measurement, IMC '07, ACM, New York,
  NY, USA, 2007, pp. 29--42.

\bibitem{ahn_sns_bfs}
Y.-Y. Ahn, S.~Han, H.~Kwak, S.~Moon, H.~Jeong, Analysis of topological
  characteristics of huge online social networking services, in: Proceedings of
  the 16th international conference on World Wide Web, WWW '07, ACM, New York,
  NY, USA, 2007, pp. 835--844.

\bibitem{mhrw_facebook}
M.~Gjoka, M.~Kurant, C.~Butts, A.~Markopoulou, Walking in facebook: A case
  study of unbiased sampling of osns, in: Proceedings of IEEE INFOCOM 2010,
  INFOCOM '10, 2010, pp. 1--9.

\bibitem{cold_ends}
Y.-X. Zhu, L.~L\"{u}, Z.~Qian-Ming, T.~Zhou, A gap in the community-size
  distribution of a large-scale social networking site, arXiv:1104.0395v1.

\bibitem{physica}
L.~{L\"{u}}, T.~Zhou, Link prediction in complex networks: A survey, Physica A
  390 (2011) 1150--1170.

\bibitem{xufeng_l}
X.~Feng, J.~Zhao, K.~Xu, Link prediction in complex networks: A clustering
  perspective, arXiv:1103.4919v1.

\bibitem{Nowell}
D.~Liben-Nowell, J.~Kleinberg, The link prediction problem for social networks,
  in: Proceedings of the twelfth international conference on Information and
  knowledge management, CIKM '03, ACM, New York, NY, USA, 2003, pp. 556--559.

\bibitem{RA}
T.~Zhou, L.~{L\"{u}}, Y.-C. Zhang, Predicting missing links via local
  information, Eur. Phys. J. B 71 (2009) 623--630.

\bibitem{Nature}
A.~Clauset, C.~Moore, M.~E.~J. Newman, Hierarchical structure and the
  prediction of missing links in networks, Nature 453 (2008) 98--101.

\bibitem{Event_network}
J.~O'Madadhain, J.~Hutchins, P.~Smyth, Prediction and ranking algorithms for
  event-based network data, SIGKDD Explor. Newsl. 7 (2005) 23--30.

\bibitem{mhrw_p2p}
D.~Stutzbach, R.~Rejaie, N.~Duffield, S.~Sen, W.~Willinger, On unbiased
  sampling for unstructured peer-to-peer networks, IEEE/ACM Trans. Netw. 17
  (2009) 377--390.

\bibitem{ff}
J.~Leskovec, C.~Faloutsos, Sampling from large graphs, in: Proceedings of the
  12th ACM SIGKDD international conference on Knowledge discovery and data
  mining, KDD '06, ACM, New York, NY, USA, 2006, pp. 631--636.

\bibitem{fs}
B.~Ribeiro, D.~Towsley, Estimating and sampling graphs with multidimensional
  random walks, in: Proceedings of the 10th annual conference on Internet
  measurement, IMC '10, ACM, New York, NY, USA, 2010, pp. 390--403.

\bibitem{estimate_size}
L.~Katzir, E.~Liberty, O.~Somekh, Estimating sizes of social networks via
  biased sampling, in: Proceedings of the 20th international conference on
  World wide web, WWW '11, ACM, New York, NY, USA, 2011, pp. 597--606.

\bibitem{diffusion_ff}
M.~D. Choudhury, Y.-R. Lin, H.~Sundaram, K.~S. Candan, L.~Xie, A.~Kelliher, How
  does the data sampling strategy impact the discovery of information diffusion
  in social media?, in: Proceedings of the 4th International AAAI Conference on
  Weblogs and Social Media, ICWSM '10, 2010, pp. 34--41.

\bibitem{AA}
L.~A. Adamic, E.~Adar, Friends and neighbors on the web, SOCIAL NETWORKS 25
  (2001) 211--230.

\bibitem{hpi}
R.~E, S.~AL, M.~DA, O.~ZN, B.~{A.-L.}, Hierarchical organization of modularity
  in metabolic networks, Science 297 (2002) 1551--1553.

\bibitem{lhn}
E.~A. Leicht, P.~Holme, M.~E.~J. Newman, Vertex similarity in networks, Phys.
  Rev. E 73~(2) (2006) 026120.

\bibitem{pa}
Y.-B. Xie, T.~Zhou, B.-H. Wang, Scale-free networks without growth, Physica
  A-statistical Mechanics and Its Applications 387 (2008) 1683--1688.

\bibitem{ff_graphovertime}
J.~Leskovec, J.~Kleinberg, C.~Faloutsos, Graphs over time: densification laws,
  shrinking diameters and possible explanations, in: Proceedings of the
  eleventh ACM SIGKDD international conference on Knowledge discovery in data
  mining, KDD '05, ACM, New York, NY, USA, 2005, pp. 177--187.

\bibitem{netscience}
M.~E.~J. Newman, Finding community structure in networks using the eigenvectors
  of matrices, Phys. Rev. E 74~(3) (2006) 036104.
\newblock \href {http://dx.doi.org/10.1103/PhysRevE.74.036104}
  {\path{doi:10.1103/PhysRevE.74.036104}}.

\bibitem{Grid}
D.~J. Watts, S.~H. Strogatz, Collective dynamics of {\lq}small-world{\rq}
  networks, Nature 393 (1998) 440--442.

\bibitem{caltech}
A.~{L. Traud}, E.~{D. Kelsic}, P.~{J. MUcha}, M.~{A. Porter}, Comparing
  community structure to characteristics in online collegiate social networks,
  arXiv:0809.0690v3.

\bibitem{email}
J.~Leskovec, K.~J.~Lang, A.~Dasgupta, M.~W.~Mahoney, Community structure in
  large networks: Natural cluster sizes and the absence of large well-defined
  clusters, Internet Mathematics 6~(1) (2009) 29--123.

\end{thebibliography}







\end{document}